\documentclass[traditabstract,referee]{aa}

\usepackage[varg]{txfonts}
\usepackage{color}
\usepackage{graphicx}
\usepackage{natbib}
\usepackage{ulem}
\usepackage{epstopdf}
\usepackage{verbatim}
\usepackage{siunitx}
\usepackage{soul}
\usepackage{multirow}
\usepackage[citecolor=blue,
            anchorcolor = blue,
            colorlinks=true,
            linkcolor = blue,
            urlcolor=blue,
            breaklinks=true]{hyperref} 
\bibpunct{(}{)}{;}{a}{}{,} 
            
\newcommand{\heii}{He II}
\newcommand{\fexxiii}{Fe XXIII}
\newcommand{\fexxiv}{Fe XXIV}
\newcommand{\fexiv}{Fe XIV}
\newcommand{\feviii}{Fe VIII}
\newcommand{\sivii}{Si VII}
\newcommand{\siiv}{Si IV}
\newcommand{\fexv}{Fe XV}
\newcommand{\fexvi}{Fe XVI}
\newcommand{\ov}{O V}

\begin{document}

\title{Multi-instrument observations of a failed flare eruption associated with MHD waves in a loop bundle}
\author{G. Nistic\`o\inst{1}\thanks{ Now at George-August-Universit\"at G\"ottingen, Germany (\email{nistico@astro.physik.uni-goettingen.de}).}  
\and 
V. Polito\inst{2} 
\and
V. M. Nakariakov\inst{1} 
\and 
G. Del~Zanna\inst{2}}

\institute{Centre for Fusion, Space and Astrophysics, Department of Physics, University of Warwick, Coventry CV4 7AL, United Kingdom  \and 
Department of Applied Mathematics and Theoretical Physics, 
           University of Cambridge, Cambridge CB3 0WA, 
           United Kingdom
          }


\abstract
{We present observations of a B7.9-class flare that occurred on the 24th January, 2015, using the Atmopsheric Imaging Assembly (AIA) of the Solar Dynamics Observatory (SDO), the EUV Imaging Spectrometer (EIS) and the X-Ray Telescope of Hinode. The flare triggers the eruption of a dense cool plasma blob as seen in AIA 171\AA,\, which is unable to completely break out and remains confined within a local bundle of active region loops. During this process, transverse oscillations of the threads are observed. The cool plasma is then observed to descend back to the chromosphere along each loop strand. At the same time, a larger diffuse co-spatial loop observed in the hot wavebands of SDO/AIA and Hinode/XRT is formed, exhibiting periodic intensity variations along its length.} 
{The formation and evolution of magnetohydrodynamic (MHD) waves depend upon the values of the local plasma parameters (e.g. density, temperature and magnetic field), which can hence be inferred by coronal seismology. In this study we aim to assess how the observed MHD modes are affected by the variation 
of density and temperature.}
{We combined analysis of EUV/X-ray imaging and spectroscopy using SDO/AIA, Hinode/EIS and XRT.}
{The transverse oscillations of the cool loop threads are interpreted in terms of vertically polarised kink oscillations. The fitting procedure provides estimates for a period of $\sim$ 3.5 to 4 min, and an amplitude of $\sim 5$ Mm. The oscillations are strongly damped showing very low quality 
factor (1.5--2), which is defined as the ratio of the damping time and the oscillation period. The weak variation of the period of the kink wave, which is estimated from the fitting analysis, is in agreement with the density variations due to the presence of the plasma blob inferred from the intensity light curve at 171\AA. The coexisting intensity oscillations along the hot loop are interpreted as a slow MHD wave with a 
period of 10 min and phase speed of approximately 436 km s$^{-1}$. Comparison between the fast and slow modes allows for the determination of the Alfv\'en speed, and consequently magnetic field values. The plasma-$\beta$ inferred from the analysis is estimated to be approximately 0.1--0.3.}
{We show that the evolution of the detected waves is determined by the temporal variations of the local plasma parameters, 
caused by the flare heating and the consequent cooling. We apply coronal seismology to both waves obtaining estimations of the background plasma parameters.}

\authorrunning{Nistic\`o et al.}
\titlerunning{A failed flare eruption associated with MHD waves}

\maketitle

\section{Introduction}
Observations of magnetohydrodynamic (MHD) waves in the solar corona provide us with an important tool for the determination of the local plasma parameters using seismology \citep[e.g.][]{Roberts1984}. 
The advent of space observatories during the last two decades has increased the richness of MHD wave phenomena found in the structured medium of the solar corona at extreme ultra violet (EUV) and X-rays wavelengths, encompassing different spatial and temporal scales.
More recently, observations at even higher resolution (1$\arcsec$, 12s cadence) 
with the Atmospheric Imaging Assembly (AIA) on board the Solar Dynamics Observatory (SDO) \citep{Lemen2012}
have substantially contributed to improving our view and knowledge of MHD modes in coronal structures. 
MHD waves have been observed as large-scale disturbances propagating through the solar disk, also known as EUV global waves \citep[e.g.][]{Patsourakos2012}; 
transverse oscillations (kink waves) of coronal loops \citep{Aschwanden1999,Nakariakov1999}, prominences \citep{Hershaw2011}; propagating and standing longitudinal oscillations  (slow waves) in loops and polar plumes \citep{DeForest1998,Wang2007, Kiddie2012,Prasad2012} and quasi periodic fast wave trains in coronal funnels \citep{Liu2012,Nistico2014}. These observations have confirmed the theory of MHD modes in a cylindrical magnetic flux tube \citep[e.g.][]{Edwin1983, Roberts1984}, which is taken as the basic model to describe dynamics of a field-aligned non-uniformity of the plasma density, typical for the corona. Kink oscillations of coronal loops have received particular interest due to their abundance in the solar corona, which allows us a systematic application of coronal seismology, and to quantify their possible contribution to coronal heating \citep[e.g.][]{Goddard2016}. Indeed, after being triggered by local coronal eruptions \citep{Zimovets2015}, the wave amplitude is observed to decay exponentially in a few cycles, which is believed to be caused by resonant absorption. 

In addition to this scenario, recent studies show the existence of a further class of kink oscillations characterised by the absence of damping \citep{Wang2012,Nistico2013, Anfinogentov2013,Anfinogentov2015}. The application of coronal seismology to kink oscillations of coronal loops has very recently received a further incentive with the discovery of a Gaussian profile at the early stage of the decay trend \citep{Hood2013}, providing a new method to uniquely determine the density contrast and the inhomogeneous layer width in coronal loops \citep{Pascoe2016}. 

A combination of EUV imaging with additional observational techniques, for example in the radio band or EUV spectroscopy,
 provides us with further constraints on the determination of the coronal plasma quantities \citep[e.g.][]{Kim2012,Kupriyanova2013,Verwichte2013}. 
Indeed, for example, parameters such as the adiabatic index $\gamma$ and the molecular weight $\mu$, which are present in the definitions of the sound and Alfv\'en speeds,  are usually assumed to standard values typical for the corona ($\gamma=5/3$ and $\mu=1.27$). \citet{VanDoorsselaere2011a} have shown that the effective adiabatic index is not equal to the assumed value of 5/3. Furthermore, observations of coexisting different MHD modes in the same coronal structure can provide compelling constraints on the determination of these plasma quantities \citep{VanDoorsselaere2011b,Zhang2015}.    

In this paper, we present a multi-instrument analysis of a flare which has been observed on the 24th January, 2015, using SDO/AIA, the EUV Imaging Spectrometer \citep[EIS;][]{Culhane97} and the  X-Ray Telescope \citep[XRT;][]{Golub2007} onboard the Hinode satellite, launched in 2006. The flare triggers an eruption of a dense and cool plasma blob, driving kink oscillations in nearby cool loop threads, and forms a diffuse hot 
loop, which in addition exhibits longitudinal oscillations of the EUV intensity interpreted as a slow magnetoacoustic wave. The aim of this study is to assess the evolution of the two MHD modes observed in association with the observed variation of density and temperature. The paper is 
structured as follows. In Section 2 we describe the observation and the instruments; in Section 3 we discuss the investigation of the 3D structure of the loop; in Section 4 and 5 we present the analysis of the kink and slow waves, respectively. Discussion and a conclusion are given in Section 6.

\section{Observations and data analysis}

The eruption that we have analysed occurred in the active region NOAA 12268 and is associated with a B7.9{-class} flare as recorded by the GOES satellite, which measured a peak in the X-ray flux at ~12:00 UT on the 24th January, 2015. 
To study the event, we used data from SDO/AIA, Hinode/EIS and XRT.

\subsection{SDO AIA}

\begin{figure*}
\centering
        \includegraphics[width=17 cm ]{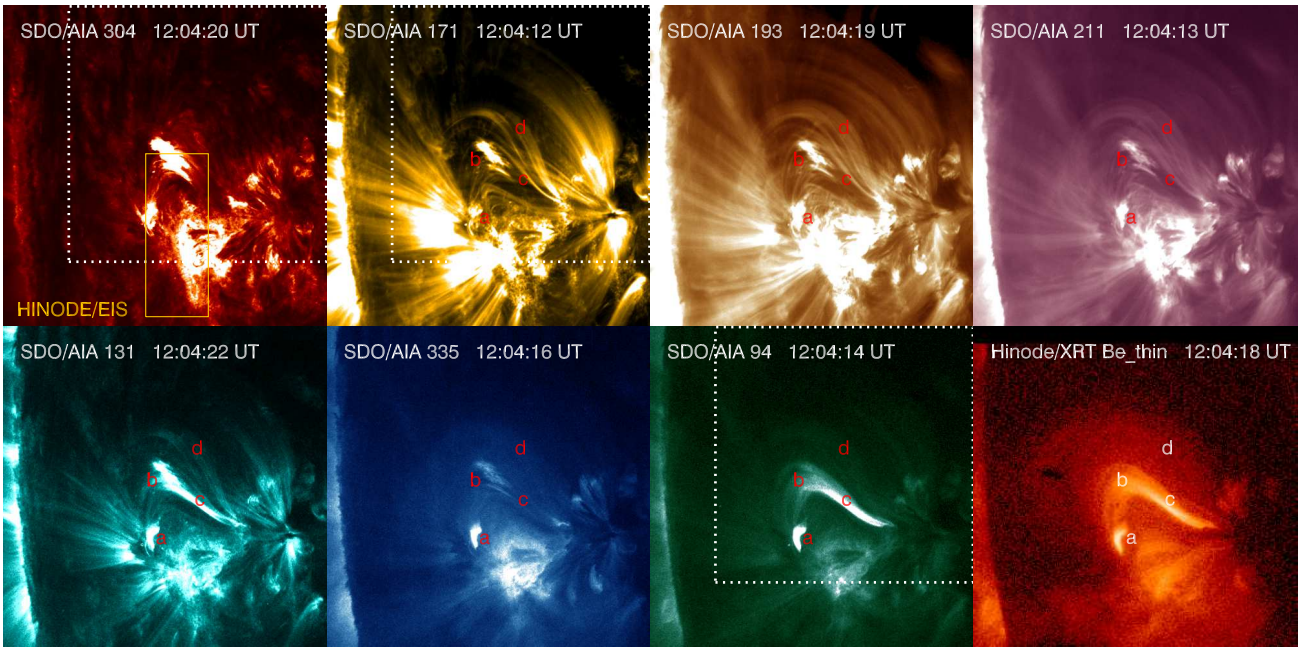}
        \caption{SDO/AIA and Hinode/XRT  images of the AR NOAA 12268 on the 24th January, 2015, at the time of the flare event. 
The box in the 304 \AA\ image shows the Hinode/EIS FOV. The labels indicate: 
{\bf a)} a small post-flare loop at the site of the initial eruption; 
{\bf b)} an expanding blob driving kink oscillations in the neighbouring loops; 
{\bf c)} a bright loop observed distinctly at 131 and 94 \AA; and
{\bf d)} overlaying unperturbed loops. The time evolution in the 171 and 94 \AA\, as well as a composite of different channels is shown in a movie available online. The composite contains the bands 304 (red), 171 (green) and 94 \AA\ (blue). The field-of-view of the movie is shown in the figure as a dotted square.}
        \label{fig1}
\end{figure*}

SDO/AIA produces full-disk images of the Sun in seven EUV wavelength bands 
(as well as in the UV) with a cadence of 12 s, a nominal pixel size of 0.6$\arcsec$ and a spatial resolution of approximately 1$\arcsec$.
We downloaded one hour of observations in all the EUV channels of AIA between 11:50 and 12:50 UT and processed 
them using the standard SolarSoft program \textit{aia\_prep.pro}.
 Figure~\ref{fig1} shows the active region observed in different AIA EUV wavebands and an X-ray image with
 the Be\_thin filter of the XRT telescope. The AIA EUV images are excellent for their resolution, but, with the exception of 
the 171~\AA\ band, are strongly multi-thermal, as described in 
\cite{odwyer_etal:10,delzanna_etal:11_aia,petkaki_etal:12,delzanna:2013_multithermal}, for example.
Assessing the plasma temperature from the AIA images is therefore prone to 
some uncertainties, and it is only by combining the AIA information with that from the 
EIS and XRT that we are able to understand the temperature evolution of this 
complex event.

\subsection{Hinode EIS }

At the time of the observations, Hinode/EIS was pointing at this active region,
even though there was a limited field of view (FoV), as shown in Fig.~\ref{fig1}. 
The EIS observations consisted of a sequence of  
`sparse' raster studies \textit{HH\_Flare\_raster\_v6} 
 from approximately 06:00 UT to 15:30 UT. Each raster
contained 20 $\times$ 2\arcsec slit steps covering an area of 60$\arcsec$$\times$ 152$\arcsec$ in $\approx$ 212 s.  The exposure time was $\approx$ 9 s.
The level 0 EIS data have been converted to level 1 data using the SolarSoft
IDL routine $eis\_prep.pro$, including standard options as described in the EIS
software notes \footnotemark[1]. In order to convert the data number (DN) to
physical units, we applied the radiometric calibration method described in
\citet{DelZanna13}, which accounts for the degradation of the detectors'
efficiency over time. EIS data were also corrected for the offset (approximately
18 pixels) between the long wavelength (LW) and short wavelength (SW) CCD channels, and the wavelength tilt along the slit \footnotemark[1].

\footnotetext[1]{http://solarb.mssl.ucl.ac.uk:8080/eiswiki/}

\begin{figure*}
\centering      
                \includegraphics[width=0.8\textwidth]{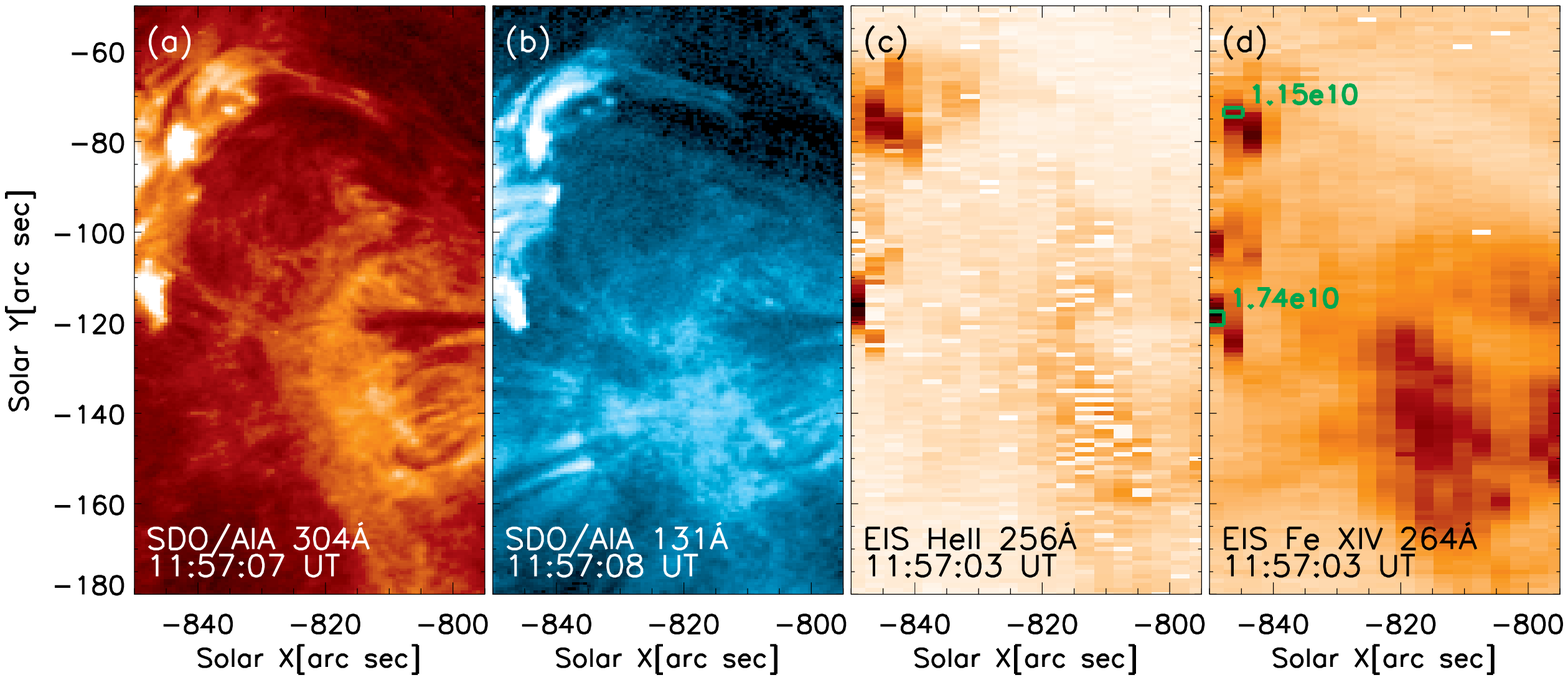} 
                \includegraphics[width=0.8\textwidth]{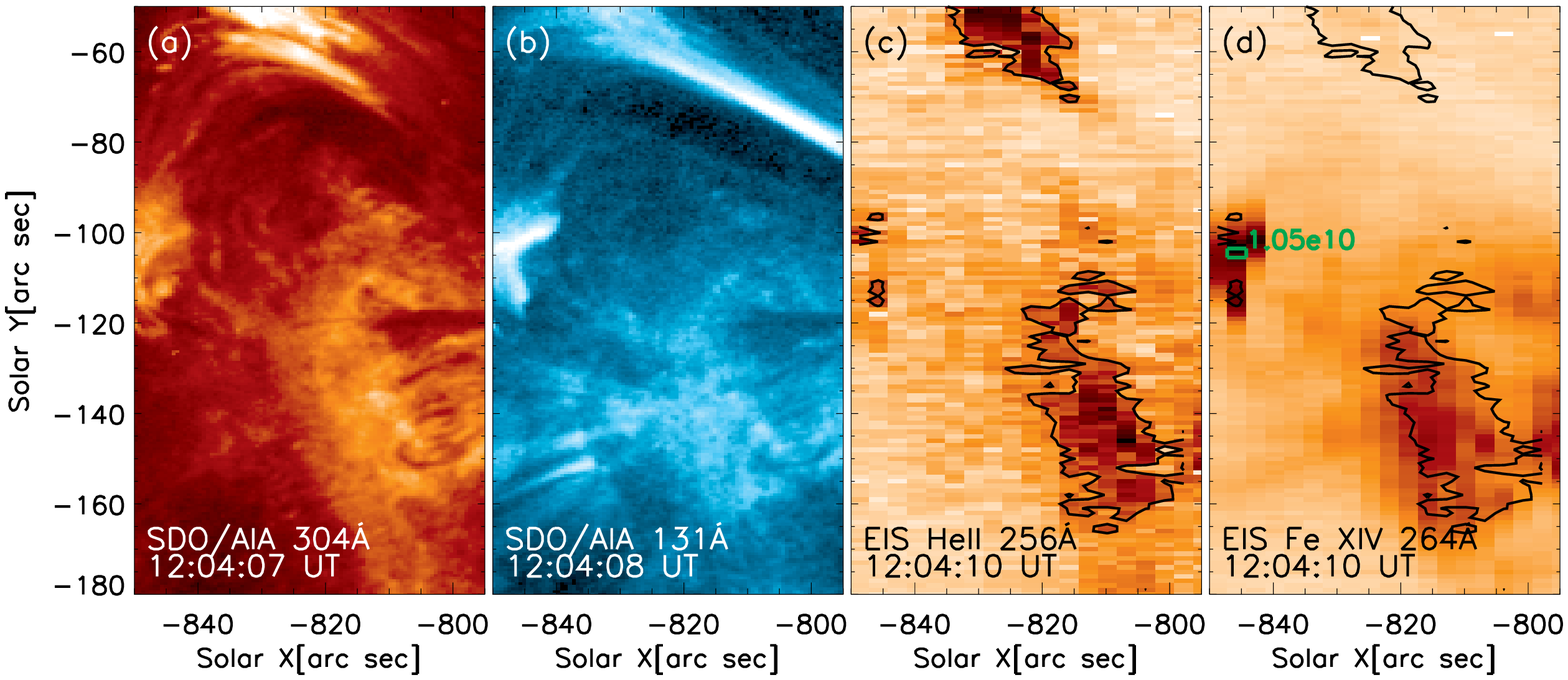} 
\caption{Images of SDO/AIA 304 (a) and 131~\AA~(b), 
Hinode/EIS \heii\ (c, T$\sim$0.05 MK) and \fexiv\ (d, T$\sim$2 MK) at 11:57 UT (top panels) and 12:04 UT (bottom panels).
The EIS intensity images are in reversed colours.}
\label{fig2}
\end{figure*}

EIS observes emission lines formed over a broad range of
temperatures, from the cool \heii\ line with a formation temperature of
$\approx$ 0.05 MK to highly ionised Fe ions (\fexxiii\--\fexxiv), which are only
visible in flaring plasma at approximately 10-20 MK. 
Fig.~\ref{fig2} shows an overview of the EIS monochromatic images in the \heii\ 256.32~\AA~(c) and \fexiv~264.79~\AA~lines (d; $T$
$\approx$ 2 MK) at two times: 11:57~UT and 12:04~UT, immediately before
and during the peak of the flare, respectively. The figures also show some context 
AIA images in the 304~\AA~filter, which is dominated by \heii\ emission, and in
the 131~\AA~filter, dominated by \feviii\ emission at 0.4 MK during non-flaring times. The intensity contours of the EIS \heii\ line are overplotted on the EIS \fexiv\ image (d) at 12:04~UT, to show the 
differences. The green squares mark the locations where the plasma density has been estimated
(values are in {cm}$^{-3}$) from the 
 intensity ratio of the EIS \fexiv\ 264.79 and 274.20~\AA~lines (see Sect. \ref{Sect:2.4}). The \fexiv\ line ratio provides
useful electron density diagnostics in the 2 MK plasma. We used atomic
calculations from the CHIANTI v.8 database \citep{delzanna_etal:2015_chianti_v8}. 
The \fexiv\ line at 264.79~\AA~is
free of significant blends \citep{DelZanna06}, whilst the \fexiv\ 274.20~\AA~is
blended with a \sivii\ line at 274.17~\AA. However, during flares, the \sivii\
line usually contributes only approximately 4 $\%$ of the intensity of the \fexiv\
274.203~\AA~line \citep{DelZanna06,Brosius13}. The \fexiv\ ratio therefore
provides a lower estimate of the electron density of the plasma.

\subsection{Hinode XRT}

XRT acquires images of the full-Sun in several X-ray broad-band filters. In this
work, we use images formed in the Be\_thin and Al\_poly filters, which are
available for the whole duration of the event under study, and which have similar
cadences ($\approx$ 2 minutes).
XRT level 0 data were converted to level 1 by using the SolarSoft
\textit{xrt\_prep.pro} routine, which removes the telescope vignetting
\citep{Kobelski2014} and subtracts the dark current from the detector's signal.
The XRT filters are broad-band and highly multi-thermal, as described in \cite{odwyer_etal:2014}.

\begin{figure*}[htpb]
       \centering
       \begin{tabular}{c c}
                \multicolumn{2}{c}{\includegraphics[width=0.8\textwidth]{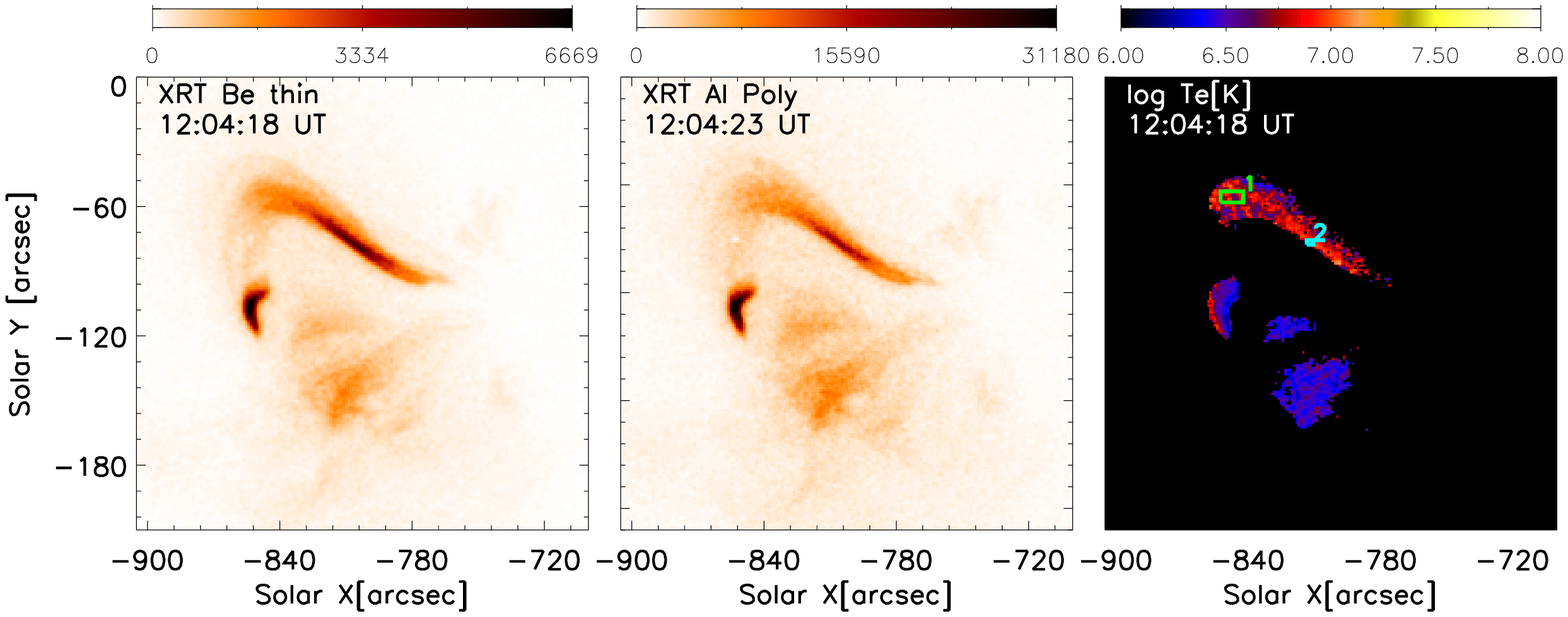}} \\
                \includegraphics[width=0.45\textwidth]{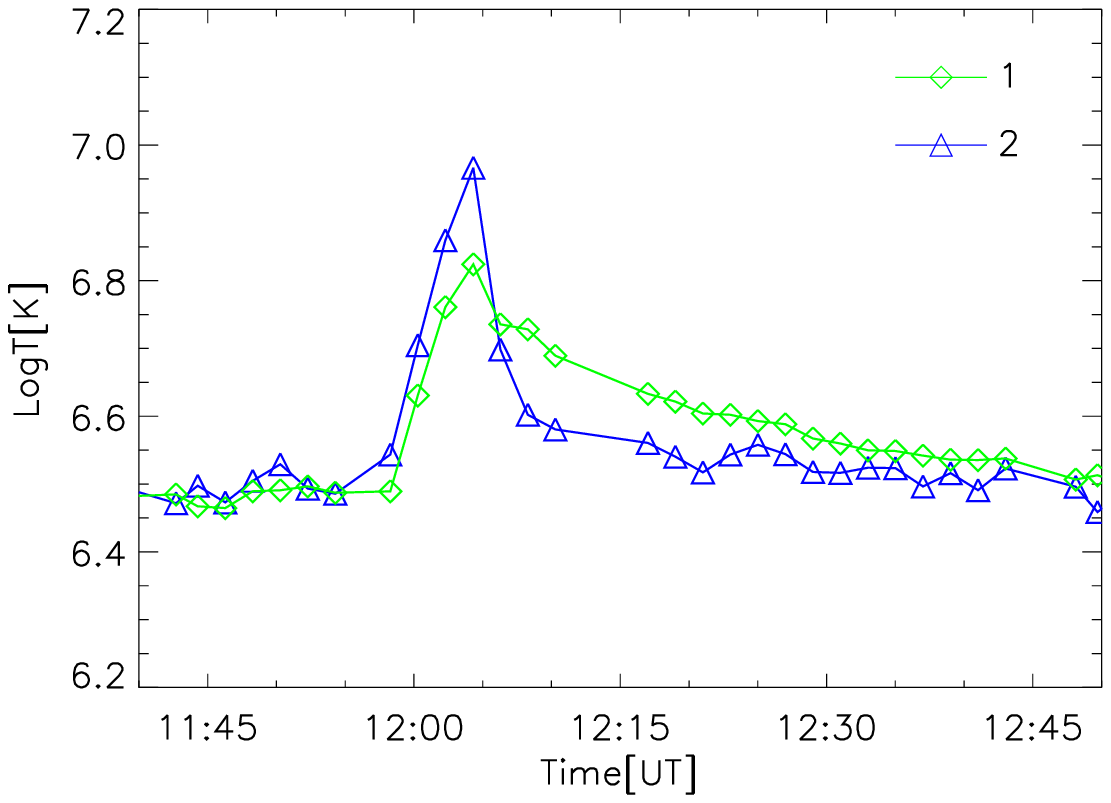} & 
                \includegraphics[width=0.45\textwidth]{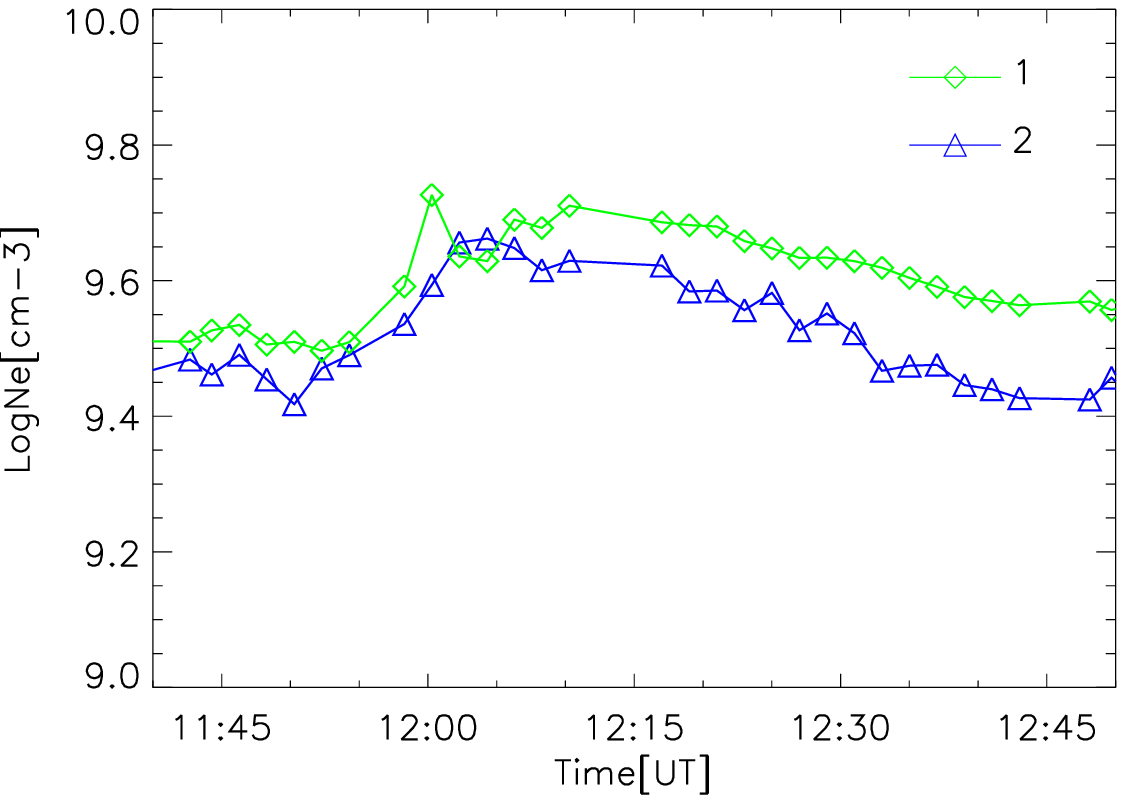} \\
            \end{tabular}
        \caption{Top: images (reversed colours) of Hinode/XRT from the Be-thin (left) and the Al-poly (middle) filters, and temperature maps obtained using the ratio of the two filters (right). The temporal evolution of the temperature map is shown in a movie available online. The entire hot loop bundle has a temperature of 6--10 MK. Bottom-left: temporal evolution of the plasma temperature at the position indicated by the boxcars 1 (green) and 2 (light blue)  in the top-right panel. Bottom-right: temporal evolution of the density in the hot loop from the boxcars 1 and 2.}
        \label{temp_maps}
\end{figure*}

For an isothermal plasma, the ratio of different XRT bands can provide
reliable temperature measurements \citep{Narukage2014,ODwyer2014}. The top-right panel of Fig.~\ref{temp_maps} shows the temperature map obtained by using the ratio of the XRT Be\_thin and
Al\_poly images at 12:04 UT, close to the peak of the flare. The temporal evolution of the temperature averaged over the boxcars 1 (green) and 2 (light blue) is given in the bottom-left panel of Fig. \ref{temp_maps}. See also the movie attached to Fig. \ref{temp_maps} to follow the evolution of the temperature map over time. The
theoretical response of each filter was obtained by convolving the filter effective area with an isothermal spectra produced by CHIANTI version 8 using chemical
abundances from \cite{Asplund09}, ionisation equilibrium calculations from
\cite{Dere2009} and a plasma density of 10$^{9}$ cm$^{-3}$. We have verified that using different values for the plasma density does not affect the theoretical XRT response function (for densities of $10^9$ and $10^{10}$ cm$^{-3}$ , the functions almost coincide), and only a much higher density can modify the observed spectrum by each XRT filter significantly.
An estimate of the plasma density for the hot loop is also obtained from the emission measure, after having defined the average temperature and a typical column depth of $\sim$5\arcsec. Figure \ref{temp_maps}-{bottom right} shows the evolution of the density averaged over the boxcars 1 and 2, respectively. The density values range between 3$\times10^9~\mathrm{cm}^{-3}$ and 6$\times10^9~\mathrm{cm}^{-3}$.

\subsection{Evolution of the plasma structures during the event}
\label{Sect:2.4}
\begin{table*}
     \centering
\caption{Timeline of the event referring to the SDO/AIA 171 images.}
        \begin{tabular}{c p{7 cm} }
        \hline
        \hline
        Time (UT)  &                 SDO/AIA                     \\
\hline
       11:51:48    & First brightenings.                       \\
       11:57:36    & Increase in the brightness in the active region.  \\       12:00:00    & Uplift of a plasma blob in 171 \AA\  and formation of a hot loop in 131 and 94 \AA.         \\
       12:02:00    & The plasma blob hits some loops and remains confined, driving kink oscillations.           \\
      ~12:11:36    & The blob diffuses along the loop threads, kink oscillations continue. Longitudinal intensity oscillations in 94 \AA.  \\
      ~12:20:00    & The cold plasma is observed to descend along the loop threads.          \\                       
\hline
\hline
        \end{tabular}
\label{tab1}
\end{table*}


%
The movie attached to Fig. \ref{fig1} shows  the evolution of plasma structures in the event, while 
Table~\ref{tab1} summarises a timeline of the event as observed in the AIA 171~\AA\ channel. 
The event is very rich in features. Indeed, a very bright small post-flare loop appears in different AIA wavebands (131, 211, 335 and 94 \AA, see feature {\bf a)}  in Fig.~\ref{fig1}).
This structure is also clearly visible in the EIS \fexiv--\fexv\ and \fexvi\ lines, indicating that the plasma is mostly emitting at approximately 2--3 MK. At the footpoint of this small flare loop, we measure densities of approximately 1.7 $\times$ 10$^{10}$ cm$^{-3}$ using the EIS~\fexiv\ line ratio (as indicated in the top panel of Fig.~\ref{fig2}), in contrast to a background density of $\sim\num{3e9}~\rm{cm}^{-3}$.  



From the region where the  small post-flare loop forms, we observe an eruption of a bright plasma blob, which is
 clearly visible in the 304 and 171 channels and may be assimilated to a small flux rope.
The bulk of this plasma has chromospheric / transition region (TR) temperatures, because it shows strong
emission in the EIS \heii\  and other TR lines.
We also note that this event was partially observed by IRIS \citep{DePontieu14}. 
 The IRIS slit-jaw images in the \siiv\ filter clearly show the eruption of this filamentary cool (T $\approx$~0.08 MK) material. Unfortunately, the oscillating loop under study was outside the IRIS FoV, and therefore we do not include any IRIS data in this work.
At the site of the eruption, the EIS \fexiv\ lines are very broad with a strong blue-shift component of approximately 100 km s$^{-1}$.

 A few minutes later (after the flare peak), at 12:04 (bottom panels of Fig. \ref{fig2}), 
this blob becomes visible only in \heii. Indeed, Fig.~\ref{fig2} shows that there is no
\fexiv\ (2 MK) emission. On the other hand, the small loop (feature {\bf a)} in Fig.~\ref{fig1}) is strongly emitting in \fexiv\, and the plasma density at the top of the loop is now $\num{1.05e10}~\rm{cm}^{-3}$ (as indicated in the bottom panel of Fig.~\ref{fig2}). 

The cool plasma expands until it hits some overlaying loops, triggering kink oscillations,
and remaining confined within them (see  {\bf b)} in Fig.~\ref{fig1}). 
These kink oscillations are only visible with AIA (they were mostly outside the IRIS and EIS FoV, see Fig.~\ref{fig1}).
Finally, this cool plasma is observed to descend back to the chromosphere 
flowing along  several loop threads in the form of coronal rain. This process lasts for approximately 20 min. 
The downflow of this cool plasma on the western side is observed in the  IRIS data and in the EIS \heii\ and TR lines (\ov, \feviii).
The apparent downflow speed along the loop threads is of the order of 10 km s$^{-1}$. The observed phenomenology would be of interest for comparison with modelling by \citet{Oliver2014,Oliver2016}, which is out of the scope of this work. 
A bright and diffuse loop in feature {\bf c)} is seen in the 131 and 94 \AA\ wavebands, while 
the larger loops observed in 171, 193 and 211~\AA\ and overlaying the active region (marked by feature {\bf d)} in Fig.~\ref{fig1})
do not show significant oscillations and are almost unperturbed. 

In contrast to the fine loop threads observed with the AIA 171 \AA,\, the sequence of the AIA 94~\AA\ images shows the formation and evolution of a well-defined diffuse hot loop, adjacent to the cool filamentary material and exhibiting longitudinal intensity oscillations (see the movie attached to Fig. \ref{fig1}). This diffuse loop appears physically distinct from the nearby loop threads observed in the \lq cooler\rq \ AIA channels. While the loop threads in 171 \AA\ are very narrow with approximately constant width ($\sim$ 1 arcsec), the \lq hot\rq \ loop is spatially more extended and approximately more uniform (we have the impression of a single and compact object), having a width of several arcsec, being thinner in the proximity of the right footpoint, larger at the apex and apparently at the left footpoint.
By combining information from AIA, EIS and XRT, we found out that this loop
has an almost isothermal temperature of approximately 8 MK. This is why the loop
is clearly visible in the AIA 94~\AA\ band, which is dominated
at this temperature by Fe XVIII \citep{delzanna:2013_multithermal}. The loop
does not emit at lower or higher temperatures, otherwise it would have been 
visible in EIS Fe XIV and Fe XXIII, for example. There is no obvious Fe XXI emission in the 
AIA 131~\AA\ band either. 
This is confirmed by the temperature analysis obtained from XRT, shown in 
Fig.~\ref{temp_maps}. We note that if only the AIA data are considered, standard inversion routines
predict instead that most of the AIA 131 emission is due to plasma at a temperature well above 10 MK.

The evolution of the loop system is accompanied by the temperature and density variations 
due to the flare, and the presence of the failed erupted plasma blob. 
In the following sections, we carefully show how these variations affect the dynamics of the transverse and longitudinal waves.

\section{3D geometry}
\label{sec_geo}
\begin{figure*}[htpb]
                \begin{tabular}{ p{6cm} p{6 cm} p{6 cm} }
                         \includegraphics[width = 6 cm]{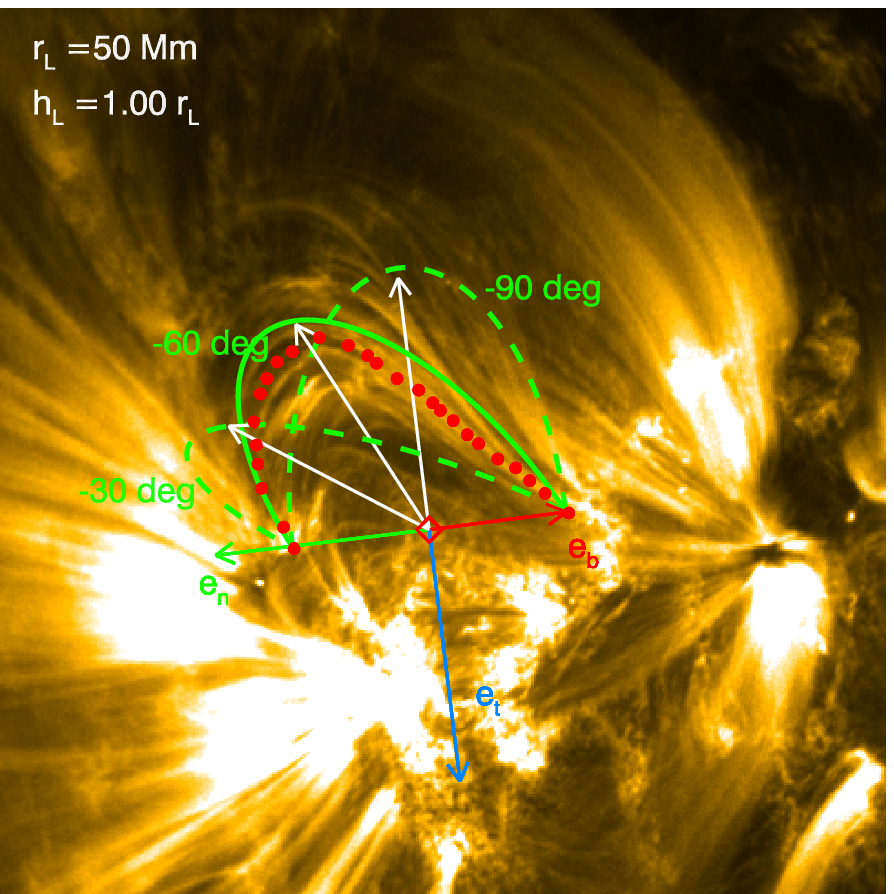} &
                        \includegraphics[width = 6 cm]{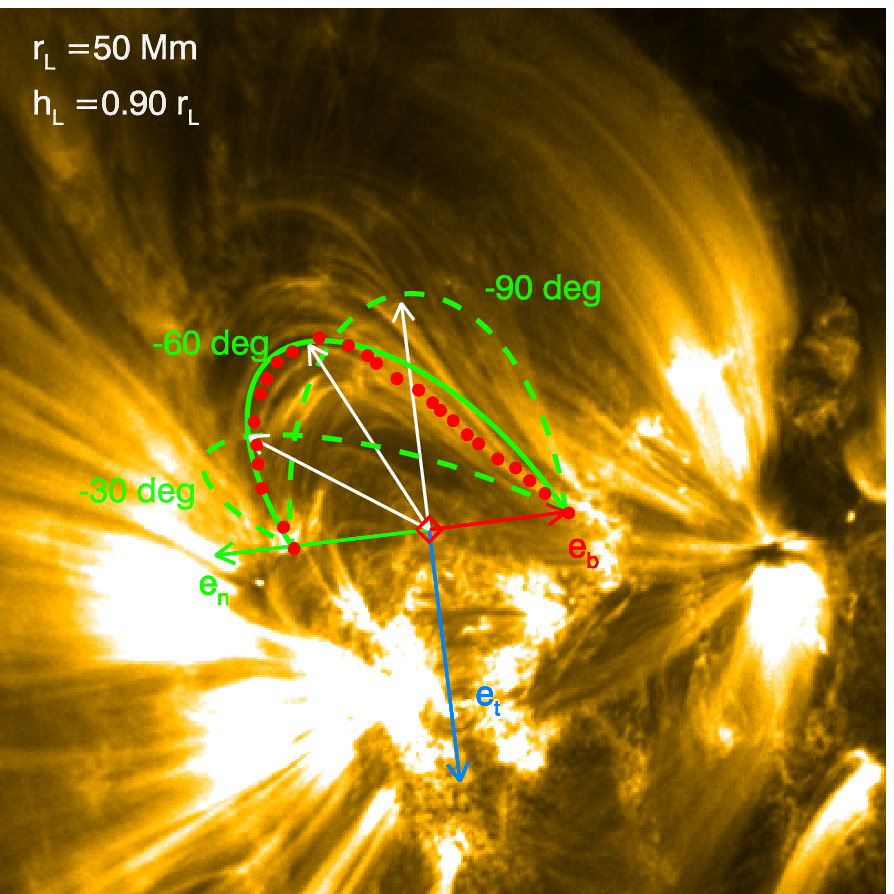} &
                        \includegraphics[width = 6 cm]{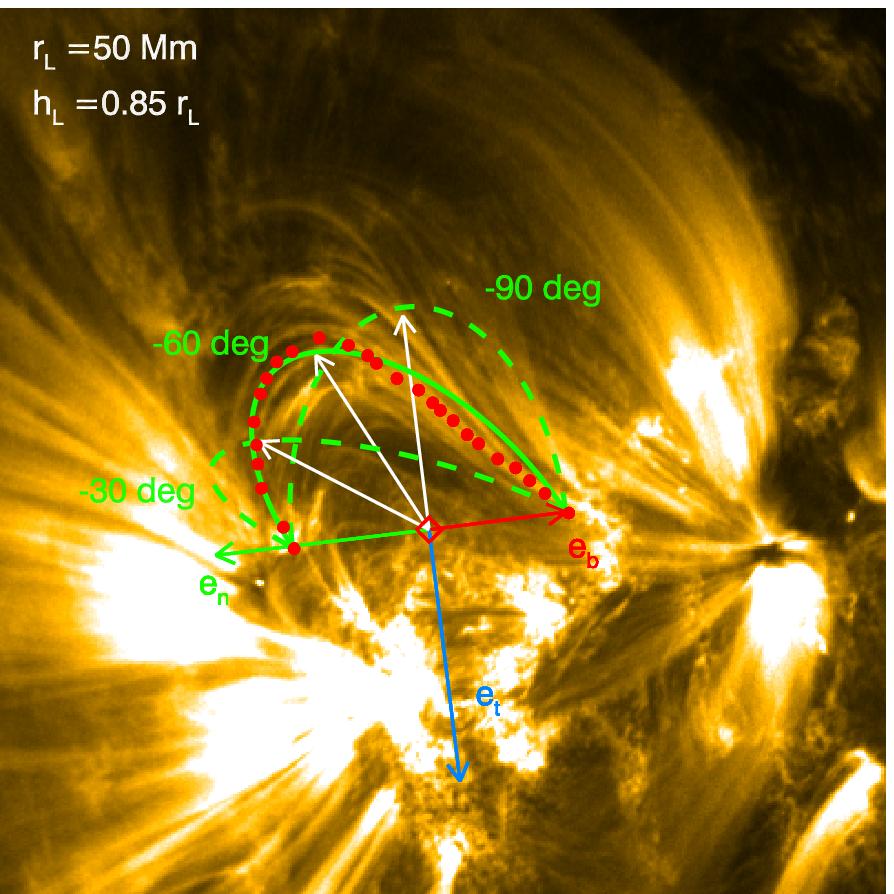} \\
                        \includegraphics[width = 6 cm]{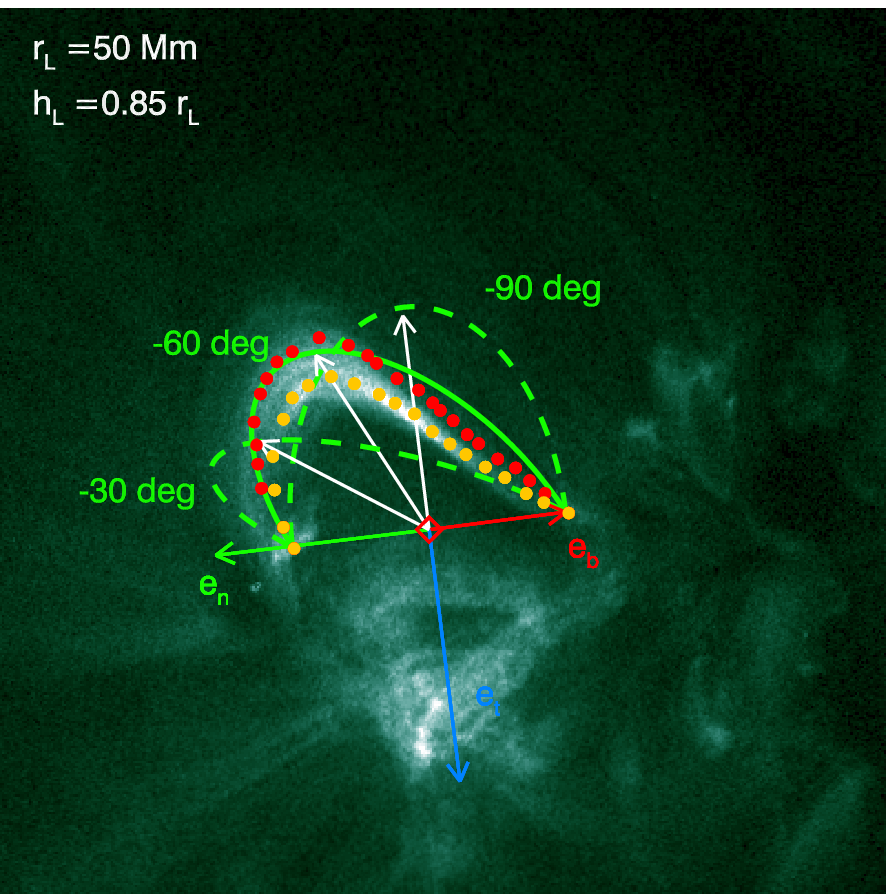} &
                                \includegraphics[width = 6 cm]{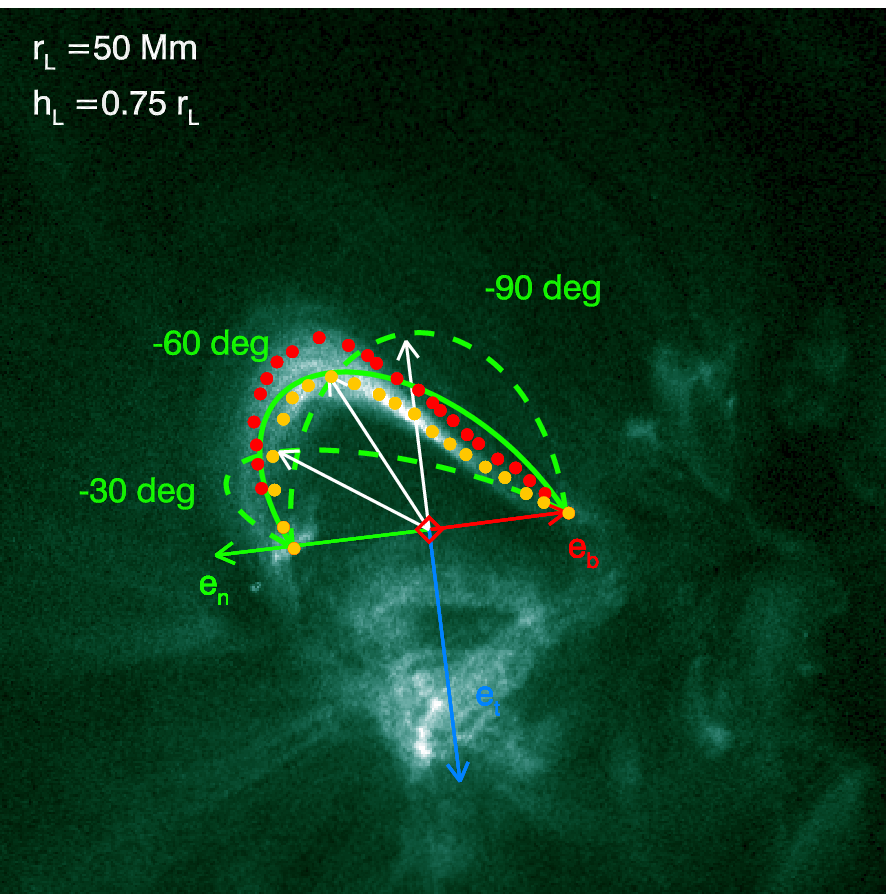} &
                        \includegraphics[width = 6 cm]{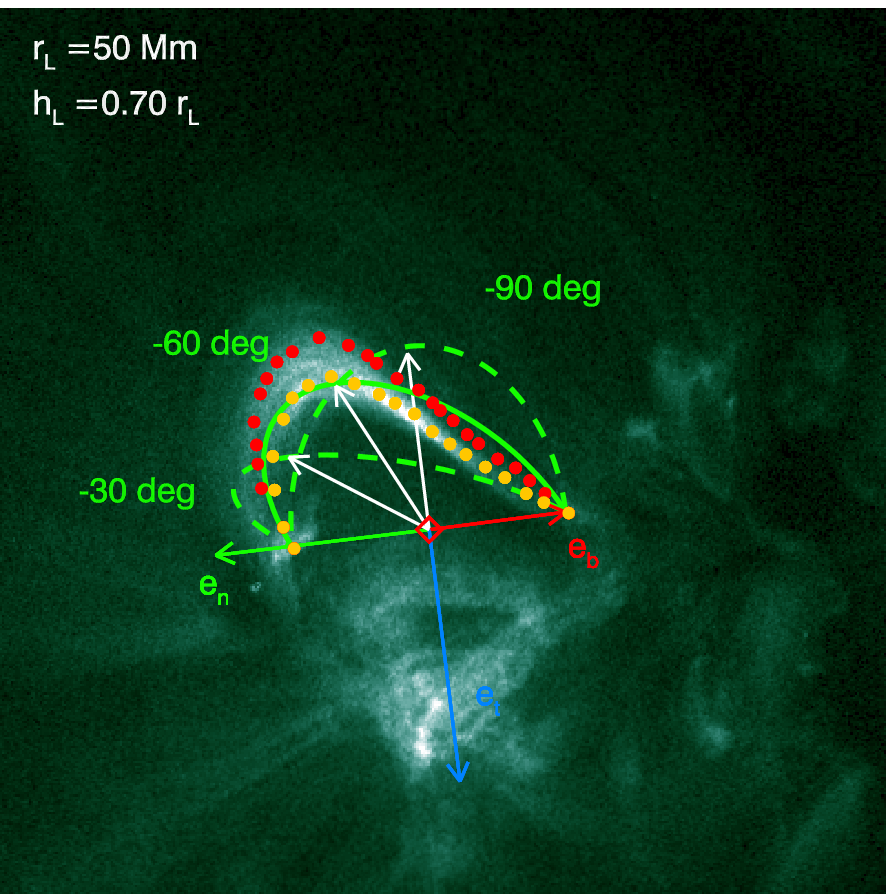} \\
\multicolumn{3}{c}{\includegraphics[width = 16 cm]{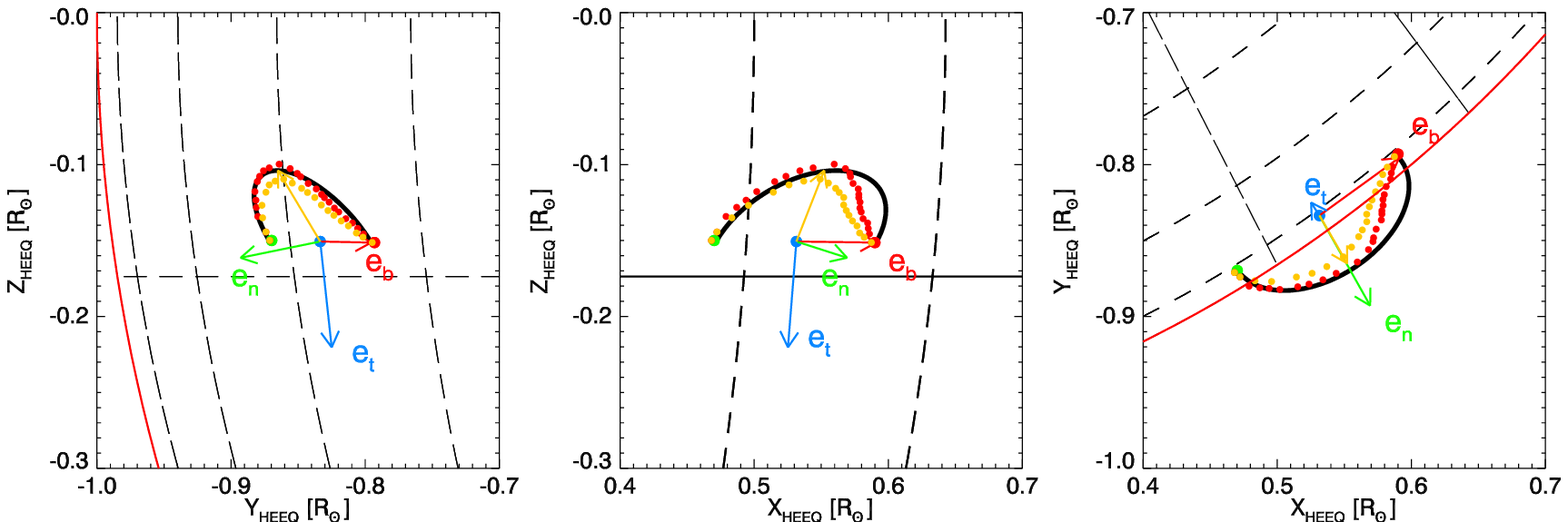}} \\
                \end{tabular}
        \caption{Three-dimensional reconstruction of the bundle of loop threads (red dots) observed with the AIA 171~\AA\ (top panels), and the hot loop (yellow dots) observed with the AIA 94~\AA\ (middle panels). The loops are best-fitted by semi-ellipses (solid green line) with an inclination angle of -60 deg with respect to the normal surface, and heights of 0.85 $r_\mathrm{L}$ in the 171~\AA, and 0.70 $r_\mathrm{L}$ in the 94~\AA\ channels. The projections of the loop system for different orientations of the HEEQ coordinate system are shown in the bottom panels.}        \label{fig3}
\end{figure*}

In the context of MHD and coronal seismology \citep{Nakariakov2016}, it is important to determine the 3D structure of coronal loops, that is, their full (and not projected) length, the inclination and azimuthal angles \citep{Nistico3d} in order to correctly interpret the periodic intensity variations, which are modulated by the periodic changes of the column depth \citep{Cooper2003}, and to unambiguously identify the polarisation of kink oscillations.
Inference of the 3D geometry by stereoscopy in this case is impossible due to the lack 
of data from the Solar Terrestrial Relations Observatory (STEREO), since the two  spacecraft  were in the back hemisphere of the Sun at the time of this observation.
 However, under some assumptions, it is possible to obtain a reliable curve that fits the series of points, which are manually determined and sample the bundle of threads observed in AIA 171 (red dots) and the diffuse loop in AIA 94 (yellow dots) (see Fig. \ref{fig3}). We adopt the technique described in \citet{Verwichte2010}.

\begin{figure*}
\centering
        \begin{tabular}{c}
                \includegraphics[width = 14 cm]{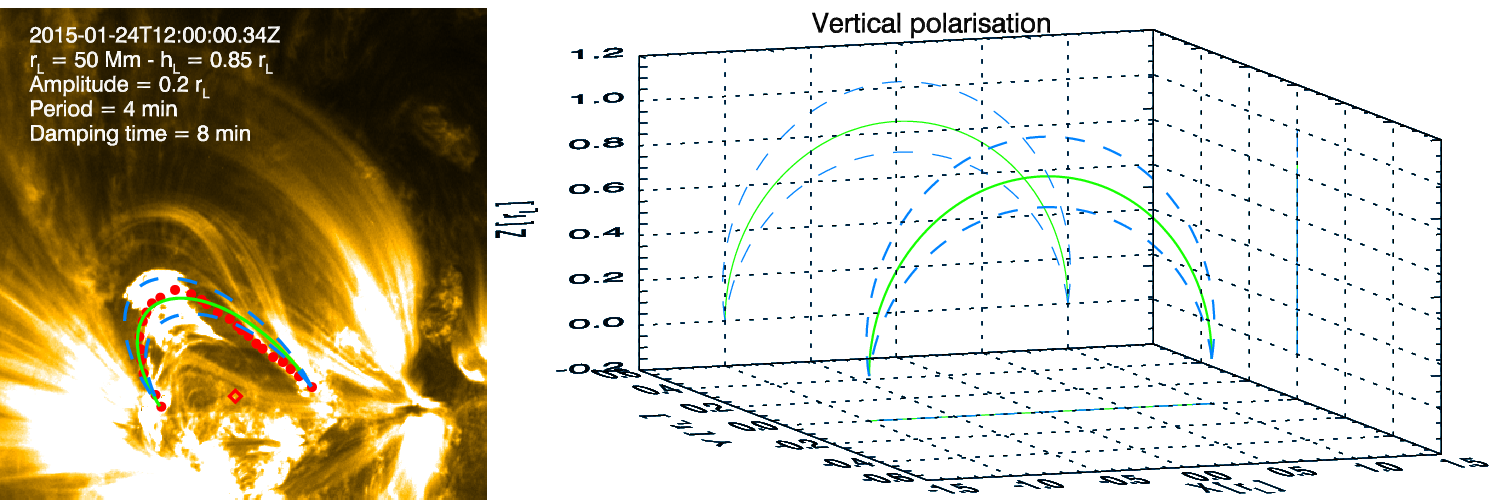} \\
                \includegraphics[width = 14 cm]{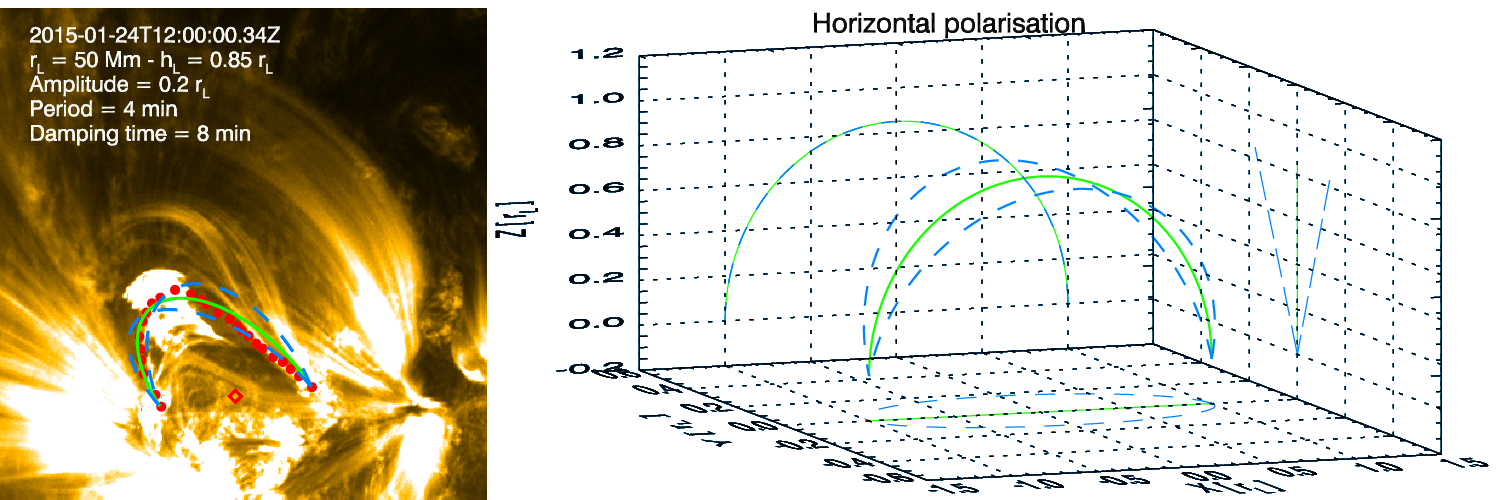} 
        \end{tabular}
\caption{Top: vertical polarisation for the transverse displacements of the coronal loop threads. The right panel shows the model for a loop at the equilibrium (green line) in a 3D Cartesian coordinate system. The distance is measured in loop radius $r_\mathrm{L}$ units. The shape of the loop at the oscillation extrema is represented by the dashed blue lines. The motion is on the $xz$ plane. 
The left panel shows the projection of the loop model and its configuration at the oscillation extrema projected in the AIA 171 FoV at the time of the start of the kink oscillations. The red dots sample a single loop thread at the equilibrium, which is fitted by the loop model in green, and the red diamond is the loop centre on the solar surface. The movement direction of the loop model at the oscillation extrema matches that of the blob and the overall loop threads.
Bottom panels: the same as described above but for the case of the horizontal polarisation mode. In this case, the motion is strictly on the $xy$ plane (right panel), and the loop configuration at the oscillation extrema does not fit the observations well (left panel). Animations of the top and bottom panels are available in the two online movies.   
}
\label{polar}
\end{figure*}

The 3D structure of the loop bundle was determined by initially
considering a semi-circle in a reference frame defined by three orthogonal axes: the loop baseline $\mathbf{e_b}$ (red arrow), 
the normal to the solar surface $\mathbf{e_n}$ (green arrow) and the vector $\mathbf{e_t}=\mathbf{e_b}\times\mathbf{e_n}$ (blue arrow). The footpoints have Stonyhurst longitude and latitude equal to (-61.6,-8.6) and (-53.4,-8.7) deg, respectively. The loop centre is consequently found as the mean between the footpoints, and has coordinates (-57.5,-8.7) deg. The footpoint half-distance represents the radius of the semi-circle, which is approximately 50 Mm. We try to match the model with the observations by varying the inclination angle $\theta$ measured with respect to the solar surface between -90 and +90 deg. The left-top panel of Fig. \ref{fig3} shows the loop as a semi-circle ($h_\mathrm{L} = r_\mathrm{L}$). For simplicity we show only three cases for $\theta$=-90 (dashed), -60 (continuous), and -30 deg (dashed green line). The white arrows represent the loop height for the different inclinations. A good approximation is obtained for $\theta=-60$ deg. This estimate is further justified by the measurements of the speed for the plasma blob, which is assumed to move along the loop plane. Indeed, as shown in the previous section, the Doppler shift, which is assumed to be related to the plasma blob expansion, defines the speed along the line-of-sight $v_\perp \approx 100$ km s$^{-1}$, while the projected speed on the plane of sky is estimated as $v_\parallel \approx 134-178$ km s$^{-1}$ (see next section). The loop inclination angle can then be found as $\theta \approx \tan^{-1}(v_\parallel/v_\perp) = 53-60$ deg. We vary the loop height to improve the fit ($h_\mathrm{L} = (1.00, 0.95, 0.90, 0.85,...) r_\mathrm{L}$, where $r_\mathrm{L}$ is the major radius). The best fit is visually found for $h_\mathrm{L}=0.85 r_\mathrm{L}$, therefore the loop seems to be slightly elliptical. However, the right loop leg is not adequately reproduced, maybe because of the presence of some further tilt in the loop plane orientation, or the departure from the plane shape, that is, a sigmoid shape. The bright and diffuse emission observed in the hot channels appears to be located slightly lower than the cooler threads, even if this statement maybe rather subjective. The loop shape in this case is sampled by the yellow dots in the central panels of Fig. \ref{fig3}. The points are fitted with a curve with a height of~$0.70 $$r_L$ {\bf(green line)}. Given the 3D orientation, we can project the series of points onto the loop plane as clearly described in the appendix of \citet{Verwichte2010}. The loop length is found by summing the distances along these points. Therefore, assuming an uncertainty of the measure of 10\%, for the cool threads observed in the 171 channel, we obtain the length $L_{171}= 141 \pm 14$ Mm, (which is comparable with that of the best-fitting ellipse, $\sim$144 Mm), while for the hot loop, the length is $L_{94} = 127 \pm 13$ Mm (the best-fitting ellipse with an height of 0.70 Mm has a length of 133 Mm). 
In addition, we have qualitatively investigated the polarisation of the transverse oscillations of the cool loop threads. Having defined the loop geometry (radius and height) and its oscillatory dynamics (amplitude, period and damping), the right panels of Fig. \ref{polar} show the model for a loop in a 3D Cartesian coordinate system at the equilibrium (green line) and at the extrema of the oscillations (blue dashed lines) for the vertical polarisation (top panel) with the motion strictly on the $xz$ plane, and the horizontal polarisation (bottom panel) where the motion takes place on the $xy$ plane. A comparison of the oscillation modes with the observations at 171 
\AA~(left panels) reveals that vertically polarised kink oscillations may match the transverse displacements of the observed loop threads (see the movies attached to Fig. \ref{polar}).  

\section{Kink oscillations of the cool threads}

\subsection{Analysis}

\begin{figure*}[htpb]
        \centering
        \begin{tabular}{c c}
                \includegraphics[width=7 cm]{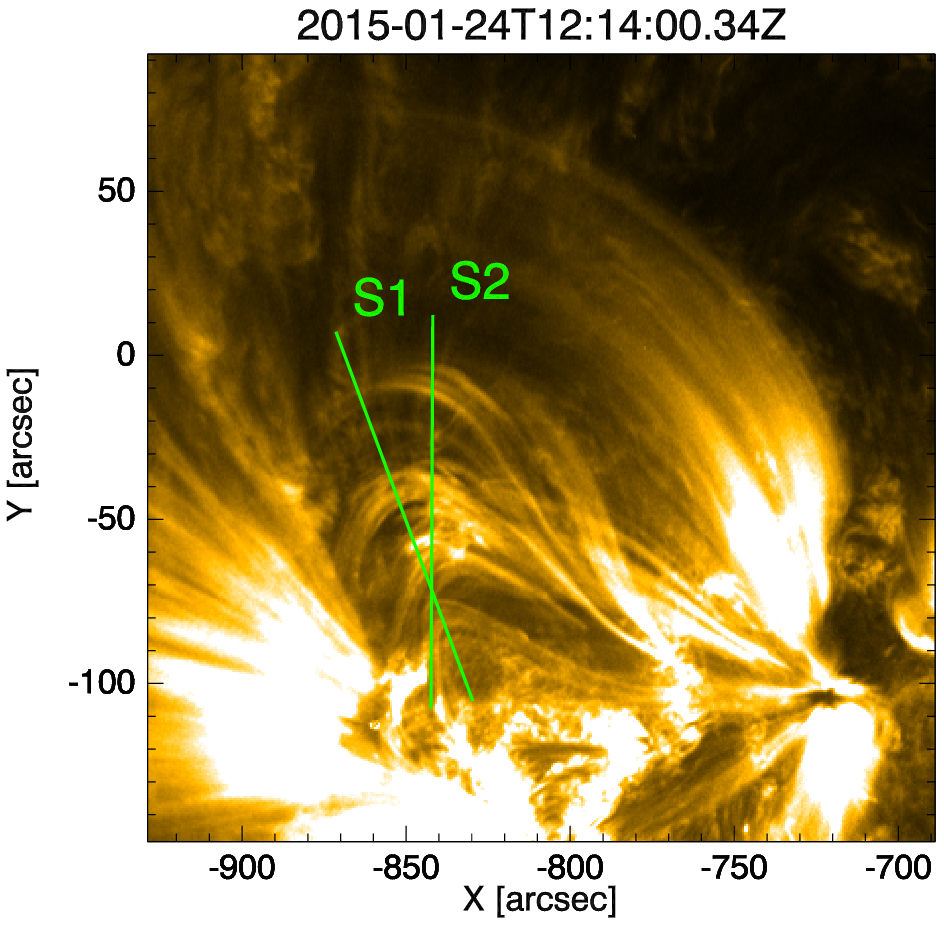} & 
                \includegraphics[width=9 cm]{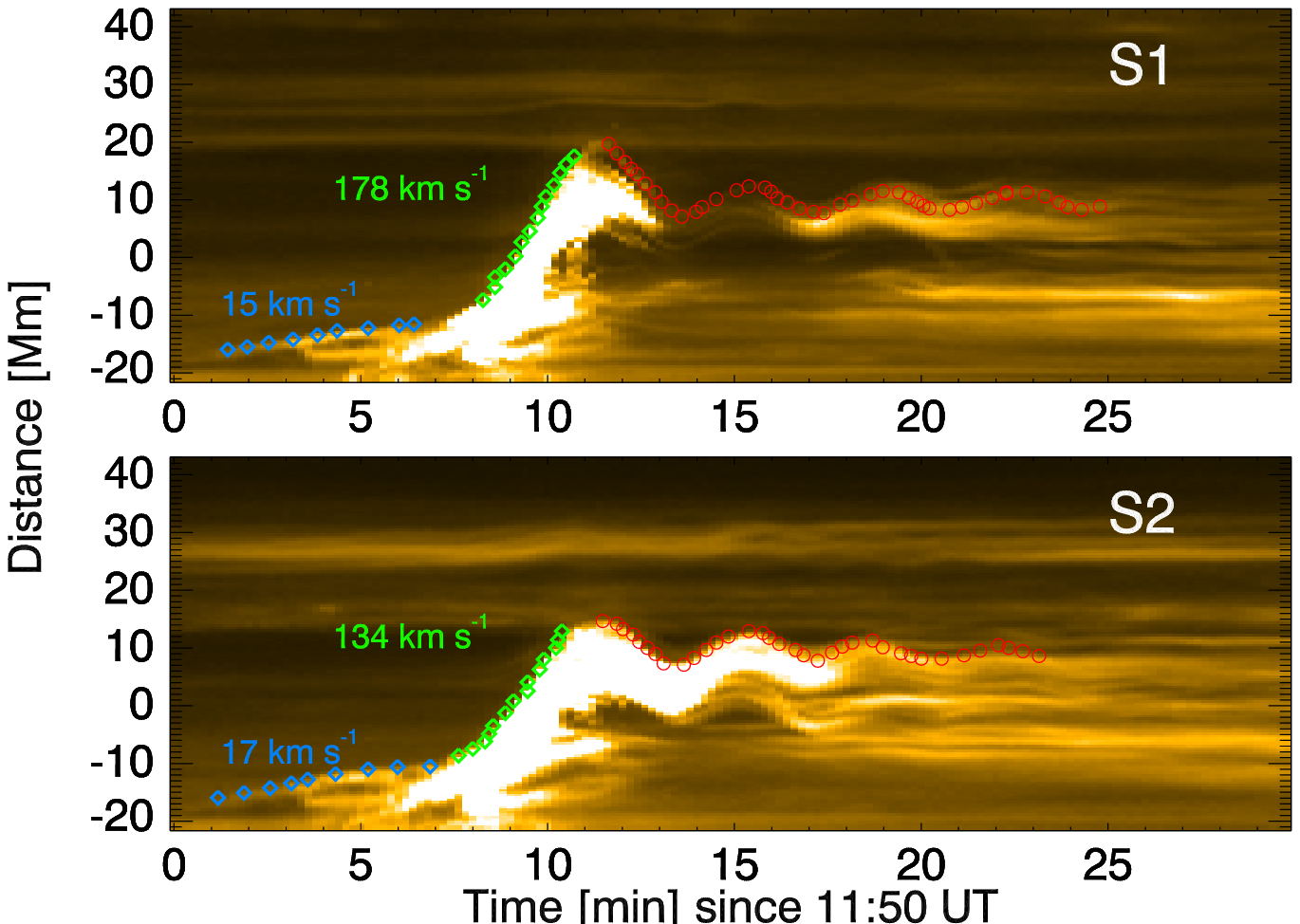} \\
        \end{tabular}
         \caption{Left: Image from SDO/AIA 171~\AA\ with the slits S1 and S2 used to make TD maps 
and trace the transverse oscillations. Right: TD maps from the S1 (top) and S2 (bottom) slits.}
         \label{slits}
\end{figure*}

To analyse the transverse oscillations of the loop threads, we have selected two slits, S1 and S2, directed as in the left panel of Fig. \ref{slits}. From these slits we have extracted the intensity for each frame of our dataset and constructed time-distance (TD) maps. The loop strands are clearly visible in the 171~\AA\ channel of AIA. The TD maps at this wavelength are given in the right panels of Fig.\ref{slits}.

In the TD maps, the signature of the expanding blob appears in the form of a very bright and inclined feature or peak. Its slope provides us with an estimate of the projected speed (green points in Fig. \ref{slits} - right panels). Before the eruption, a much slower expansion with a linear speed of 15--17 km s$^{-1}$ is measured (blue points). After this phase, the blob is {expanding} with a projected velocity of approximately 178 km s$^{-1}$ along slit S1, while in S2 the speed is lower, 134 km s$^{-1}$, since this direction does not exactly match the one of the expanding blob. The expansion of the blob displaces the loop threads from their equilibrium, which undergo transverse oscillations (red points in the TD maps). The patterns in the TD maps are composed of several strands, which are not very easy to track, and oscillate collectively. The oscillations in the right panels of Fig. \ref{slits} are tracked by following the upper rim of the oscillating bundle by eye. Therefore, we characterise the oscillatory dynamics of the overall bundle rather than that of each single loop thread.

\begin{figure*}[htpb]
  \centering
  \begin{tabular}{c c c}
        \includegraphics[width=5 cm]{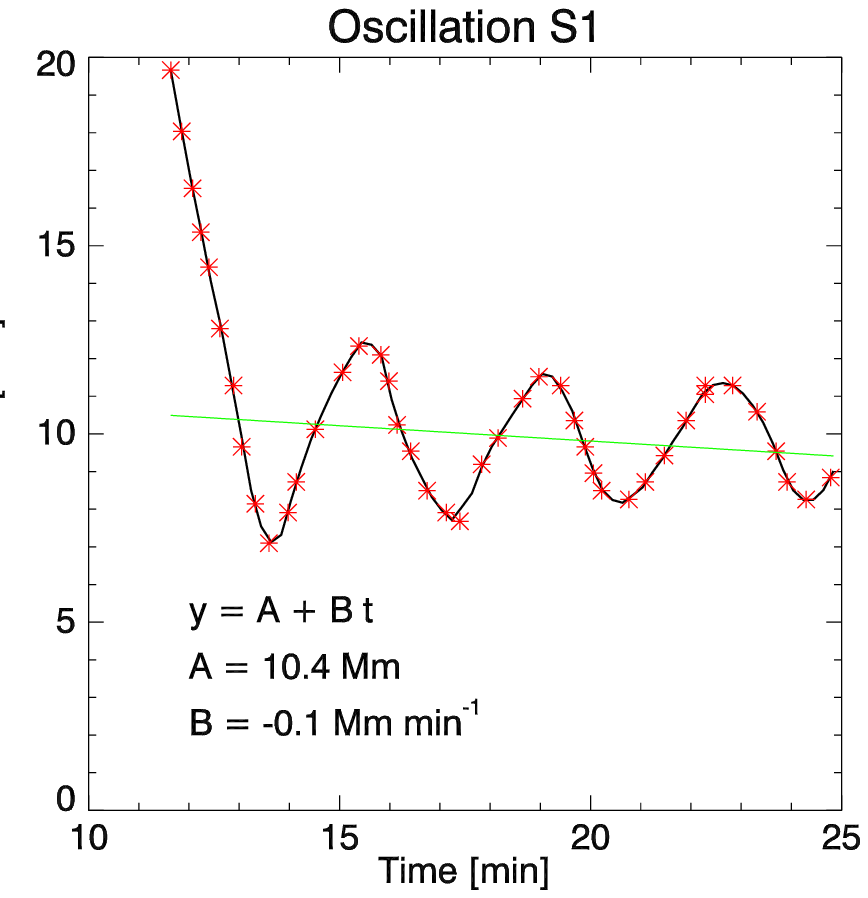} &
        \includegraphics[width=5 cm]{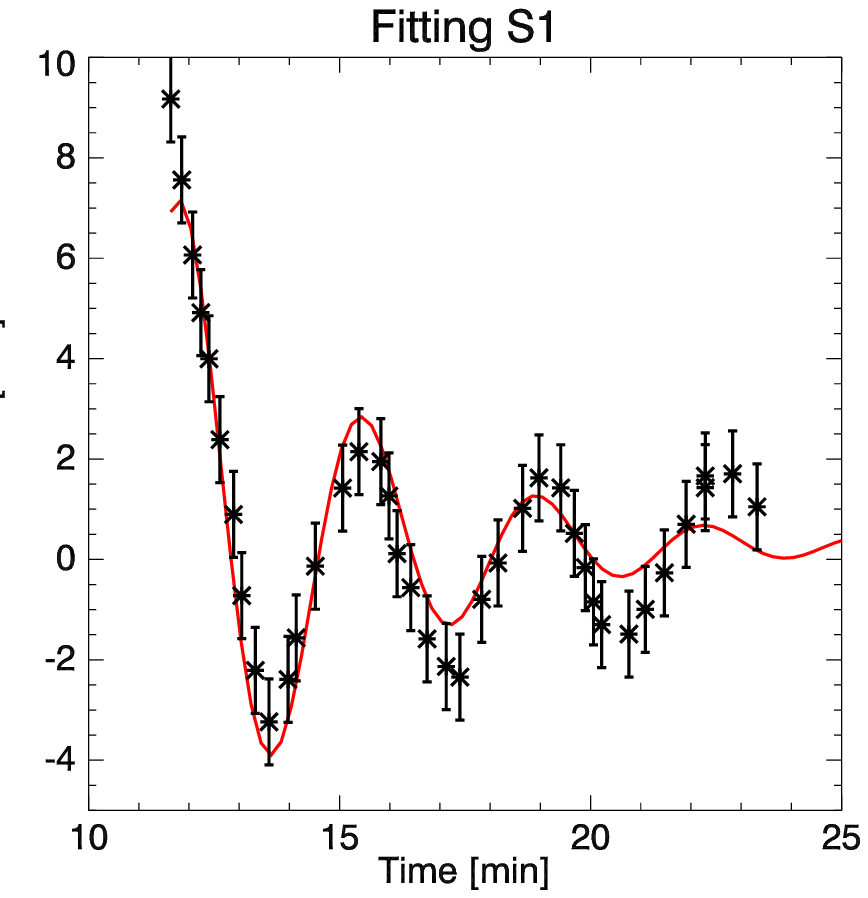} & 
        \includegraphics[width=5 cm]{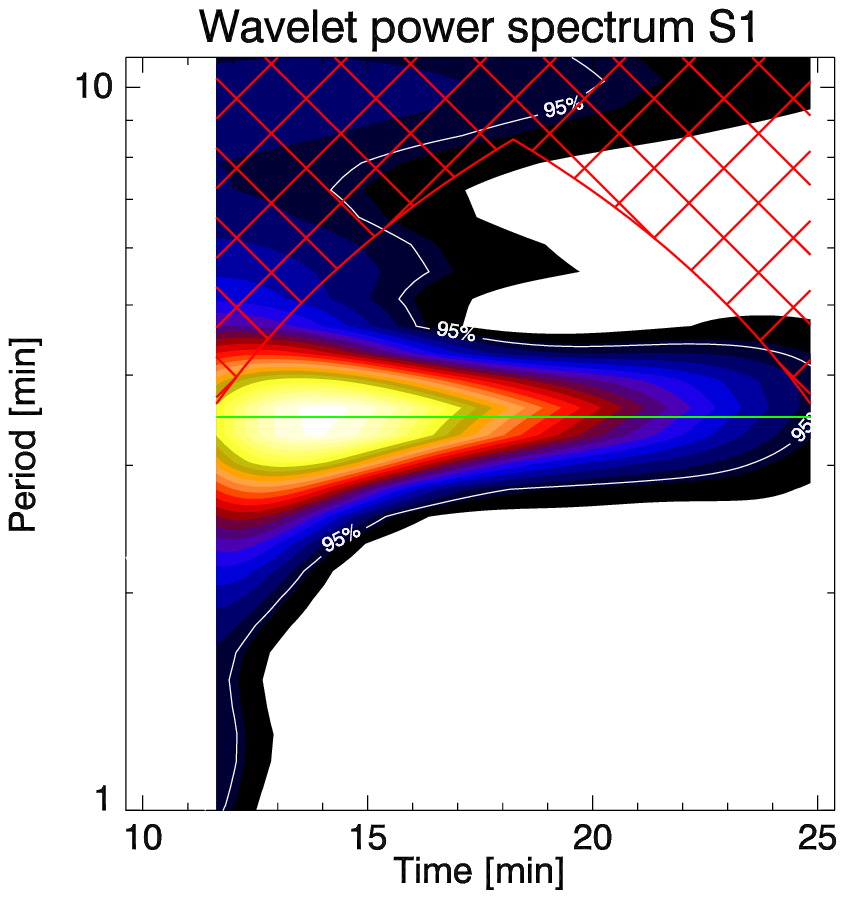} \\
        \includegraphics[width=5 cm]{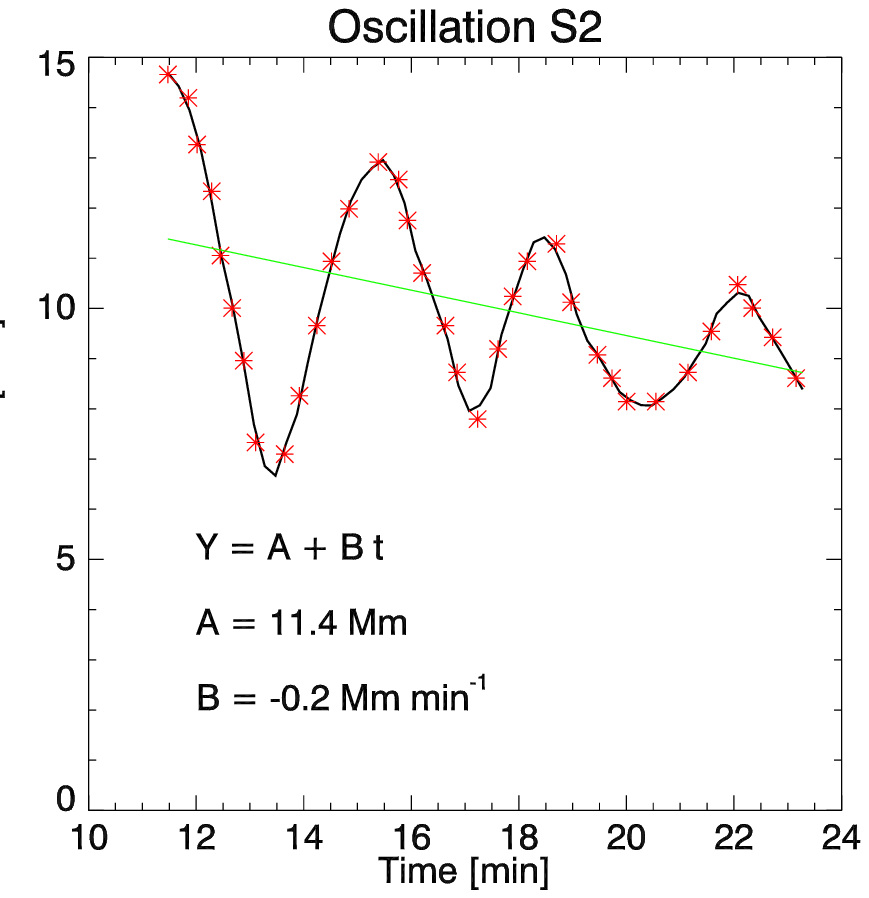} &
        \includegraphics[width=5 cm]{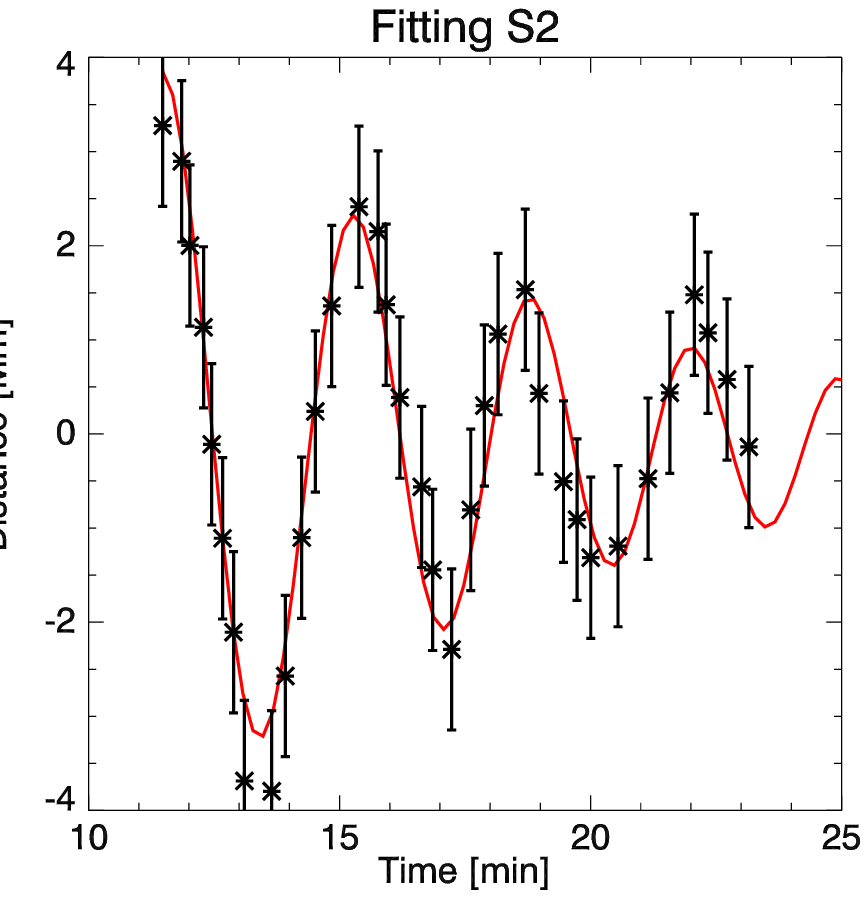} &
        \includegraphics[width=5 cm]{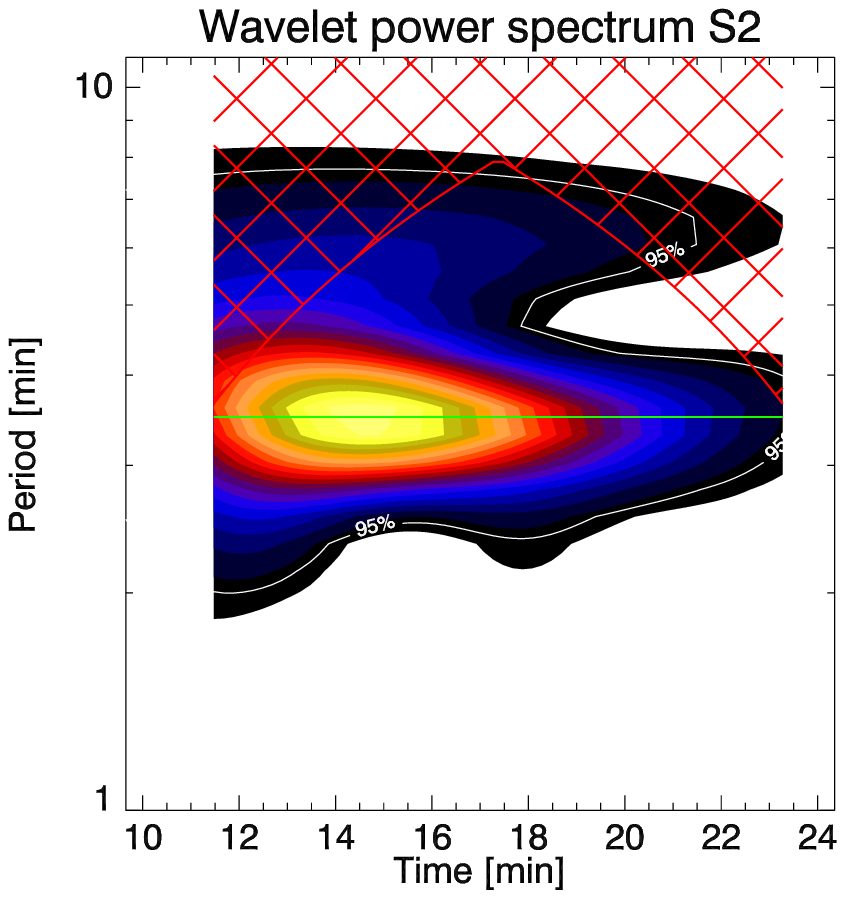} \\
  \end{tabular}
   \caption{Time series of the oscillations S1 and S2 (left),  wavelet power spectra of the time series (centre), 
and plots of the period vs time (right). The period is between 3.7 and 4 min. 
There is a very small variation of the period from the fitting analysis, however the green lines in the wavelet power spectra have a null slope.}
   \label{kink}
\end{figure*}

\begin{table*}[htpb]
   \centering
  \caption{List of the fitting parameters for the observed oscillations.}
  \begin{tabular}{ c | c c c c c c c }
\hline
\hline
Oscillation  &     $t_0$  & $y_0$     &   $A$    &  $P$  &  $P'$                     &    $\phi$ & $\tau$ \\
                  &     [min]    &  [Mm]       &   [Mm]  &  [min] &  [min min$^{-1}$]  &   [deg]    &   [min] \\
\hline
     S1      & 11.6 & 0.3$\pm$0.1        & 7.3$\pm$0.5~             & 3.7$\pm$0.2 &-0.02$\pm$0.03& -24$\pm$8 & 3.7$\pm$0.5 \\  
     S2      & 11.5 & -0.1$\pm$0.2       & 4.0$\pm$0.5~             & 4.0$\pm$0.3 &-0.05$\pm$0.03& -2$\pm$11 & 7.9$\pm$2.0 \\ 
     S3       & 10.1 & 0.02$\pm$0.01   & 0.4$\pm$0.1$^\ast$ & 9.7$\pm$0.4 &0.05$\pm$0.02 & -110$\pm$8  & 40.3$\pm$14.7\\
\hline
\hline
  \end{tabular}

\tablefoot{Kink oscillations S1 and S2 are shown in Fig. \ref{kink}, while the longitudinal oscillation S3 refers to Fig. \ref{94img}. $^\ast$The amplitude of the slow MHD wave S3 is not in Mm units but normalised to the loop length.}
  \label{table1}
\end{table*}

The oscillatory patterns have been fitted using the MPFIT routines \citep{Markwardt2009} with the following function
\begin{equation}
 y(t) = y_0 + A \cos\left(\frac{2 \pi (t-t_0)}{P + P' (t-t_0) } + \phi\right) \exp\left(-\frac{t-t_0}{\tau}\right),
\label{eq_fitting} 
\end{equation}
which assumes a priori a linear dependence of the period on the time, with $P'$ being a variation rate of the period. Before applying the fitting routine, the time series was detrended with a background linear fit.
The fittings were weighted by the errors of each data point, which was taken to be approximately 2 pixels ($\sim$ 1 Mm). The amplitudes of the oscillations (Table \ref{table1}) are estimated to be approximately 4--6 Mm. The initial period $P$ of the oscillations is between 3.7 and 4.0 min. The period rate change $P'$ is negative in both cases (even if the standard deviations associated to these estimates make them relatively insignificant), and is consistent with a decrease of the density, as we will see in more detail in the next subsection.  
The wavelet power spectra of the time series in the central panels of Fig. \ref{kink} 
shows that the period of the kink oscillations is approximately 3.5 min (green line).

\subsection{Coronal seismology with kink waves}

The transverse displacements observed in the loop are interpreted in terms of the fundamental standing fast magnetoacoustic kink wave. The phase speed $V^{(\mathrm{K})}_\mathrm{ph}$ 
is determined by the loop length and the period of the oscillation, that is,
\begin{equation}
 V^{(\mathrm{K})}_\mathrm{ph}  = \frac{2L}{P}.
\end{equation}

In this case, given the length $L=L_{171}=141\pm1$4 Mm and the period of $P=3.9\pm0.3$ min, the phase speed is $V^{(\mathrm{K})}_{\mathrm{ph}} = 1205\pm152~\rm{km~s}^{-1}$ (see also Eqs. \ref{app_ph_1}-\ref{app_ph_2} in Appendix \ref{appendix}). From theoretical modelling of MHD modes in a plasma cylinder \citep[e.g.][]{Edwin1983}, the phase speed $V^{(\mathrm{K})}_\mathrm{ph}$ for long-wavelength kink oscillations (in comparison with the minor radius of the loop) is the kink speed $C_\mathrm{K}$, which is the density-weighted average of the Alfv\'en speeds inside and outside the oscillating plasma cylinder:
\begin{equation}
    C_\mathrm{K} = \left( \frac{\rho_0 C_\mathrm{A}^2 + \rho_\mathrm{e} C_\mathrm{Ae}^2}{\rho_0+\rho_\mathrm{e} }\right)^{1/2}.
\end{equation}
In the low-$\beta$ plasma regime, typical for coronal active regions, the expression above can be approximated as
\begin{equation}
   C_\mathrm{K} \approx C_\mathrm{A} \left( \frac{2}{1+\rho_\mathrm{e}/\rho_0} \right)^{1/2},
   \label{ck}
\end{equation}
where $C_\mathrm{A} = B/\sqrt{4 \pi \rho_0}$ is the Alfv\'en speed, and $\rho_0$ is related to the number density $n_0$ with the relation $\rho = \mu m_p n_0$.
Therefore, changes in density can affect the values of $C_\mathrm{K}$, and consequently the period $P$, as also found in the decayless oscillation event described by \citet{Nistico2013}. Indeed, using the expressions above it is relatively straightforward to show that
\begin{equation}
 P = \frac{L}{B}\sqrt{8\pi\rho_0(1+\rho_\mathrm{e}/\rho_0)}.
\label{P_vs_n}
\end{equation}

\begin{figure}[htpb]
  \centering
        \includegraphics[width=7 cm]{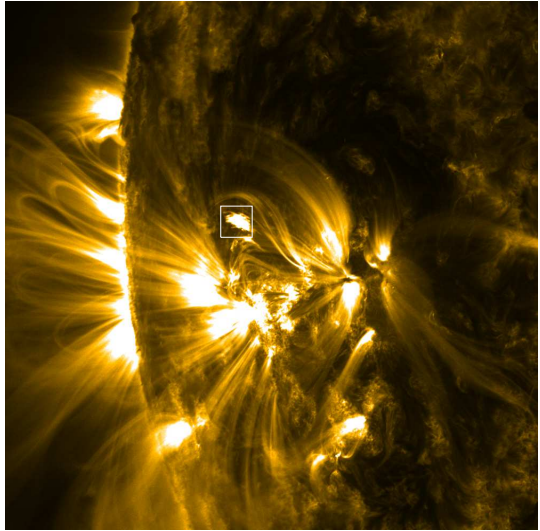} 
        \includegraphics[width=7 cm]{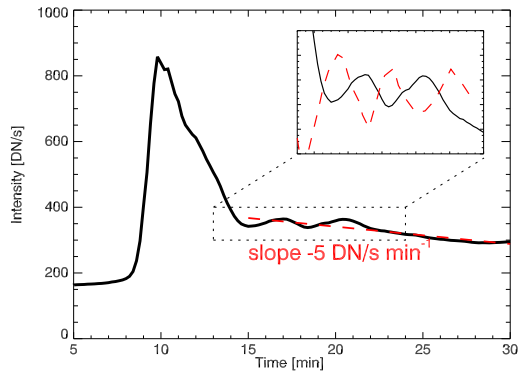} 
   \caption{Intensity time series in the AIA 171~\AA\ band (bottom) starting at 11:50 UT, 
obtained from a boxcar centred on the cool plasma blob (top). 
The intensity shows a strong peak at the flare peak time, and then it drops very quickly in approximately 5 min. In the inset plot we show that the oscillations in the intensity profile (black) are in antiphase with the loop thread displacement (dashed red line).}
   \label{intensity}
\end{figure}

After its expansion, the blob is observed to slowly diffuse and descend along the loop threads. The observed longitudinal flow could also be induced by the ponderomotive force associated with the nonlinear kink oscillation \citep[e.g.][]{Terradas2004}, but the theory of this effect is not elaborated enough to make any quantitative comparison. Perhaps the weakly compressive nature of long-wavelength kink oscillations would allow one to adopt the results obtained for the ponderomotive force in  Alfv\'en waves \citep[e.g.][]{Tikhonchuk1995,Verwichte1999,Thurgood2013}. To investigate the influence of the downflowing plasma blob on the period of the kink oscillations, we analyse the intensity time series from the SDO/AIA 171~\AA\ band.
Figure \ref{intensity} shows the intensity time series $I$ averaged over a boxcar centred on the cool blob (bottom panel). 
The profile resembles that of a shock with a very sharp ramp (at $\sim$ 9 min), an overshoot (after 10 min)
and a weak decaying tail (after $\sim$15 min). 
Since $I \propto n_0^2$, then variations of the intensity on time-scales larger than the kink period can be expressed as $\delta I/ I = 2 \delta n_0/n_0$, assuming that the plasma does not dramatically change its temperature, and the angle between the local loop segment and line-of-sight remains constant, as the observations suggest. 

On the other hand, from Eq. \ref{P_vs_n}, the variations of the period with respect to the density 
(considering the inner density $\rho_0$ or equivalently $n_0$) are $\delta P/P = \delta n_0 /2n_0$. 
Therefore,  $P$ changes with respect to $I$ as $\delta P/P = \delta I /4I$. For $\delta I = -50$ DN 
in a time interval, $\Delta t = $ 10 min and $I=350$, $\delta P/P \approx - 0.03$. By taking $P=3.9$ min, it is easy to show that $\delta P \approx -0.03 P =-0.11$ min, and therefore  $P' \approx \delta P/\Delta t = -0.01$ min min$^{-1}$, which is consistent with the values obtained from the fittings.
It is worth mentioning that the oscillations that are seen in the intensity profile (see the inset plot in Fig. \ref{intensity}-bottom panel) are in anti-phase with the displacement of the loop threads. This can be explained in terms of vertically polarised oscillations of a bundle of loop threads as suggested by \citet{Aschwanden2011}. In general, such variations are determined by the periodic change in the column depth \citep{Cooper2003} and the correct polarisation mode can be inferred by appropriate forward modeling, as discussed in \citet{Verwichte2009} and recently shown by \citet{Ding2016a,Ding2016b}.  
We can now estimate the Alfv\'en speed and the magnetic field (see Eqs. \ref{eq:ca}--\ref{eq:sigma_b}).
We considered the following values: $L=L_{171}= 141\pm14$ Mm, $P = 3.9\pm0.3~\rm{min}$, $\mu\approx 1.27$, $m_p=\num{1.67e-24}$ g, $n_0 =(1.0\pm0.5)\times 10^{10}$ cm$^{-3}$ and $n_e = (3.0\pm1.5)\times 10^9$ cm$^{-3}$ (we have assumed uncertainties for the densities of 50\%). We note that the density of the cool threads is very difficult to measure. Because of its lower spatial resolution with respect to SDO/AIA, Hinode/EIS observed the loop as a single thread only. The density estimate for $n_0$ has been obtained from the emission measure, assuming a column depth of 3$\arcsec$, photospheric abundances and volume filled.
We obtain an Alfv\'en speed $C_\mathrm{A}=  972 \pm 146$ km s$^{-1}$, and a magnetic field of $B = 50 \pm 12$ G.
A summary of the coronal parameter estimates is given in Table \ref{table2}.
\begin{table*}[htpb]
   \centering
\caption{List of the physical parameters for the observed fast and slow MHD waves.} 
  \begin{tabular}{ c | c c c c c c c c c}
\hline
\hline
Wave &  $L$        & $P$      & $T$ & $n_e$            & $n_0$   & $V_\mathrm{ph}$ & $C_\mathrm{S}$ & $C_\mathrm{A}$ & $B$\\
     & [Mm]        &[min]     & [MK]&[10$^9$ cm$^{-3}$]&[10$^9$ cm$^{-3}$]& [km s$^{-1}$]   & [km s$^{-1}$]  &   [km s$^{-1}$]& [Gauss] \\
\hline
 Kink              & 141$\pm$14 & 3.9$\pm$0.3 &  -                   &  3.0 & 10.0&1205$\pm$152 & -                    & 972$\pm$146   & 50$\pm$12 \\ 
 \hline 
 Slow-sound  & \multirow{2}{*}{127$\pm$13} & \multirow{2}{*}{9.7$\pm$0.4} & 8.0$\pm$2.0 &  -     &  -     & \multirow{2}{*}{436$\pm$47}    & 416$\pm$52 & -                        &      -              \\
 Slow-tube    &  &  &10.0$\pm$2.5& -      &  6.0 &   & 465$\pm$58 & 1255$\pm$207 & 50$\pm$12 \\   
\hline
\hline
  \end{tabular}
  
\tablefoot{The loop length $L$, the period $P$, the temperature $T$, the densities $n_e$ and $n_0$ are estimated from the observations and are used as input values for the determination of the sound and Alfv\'en speeds. The magnetic field is estimated via coronal seismology. In the slow-tube wave approximation, the inner density $n_0$ of the hot loop is found from the Alfv\'en speed and the value of magnetic field, which has been previously inferred from the kink wave.}

  \label{table2}
\end{table*}

\section{Longitudinal oscillations in the hot loop}

  \subsection{Analysis}

\begin{figure}[htpb]
\centering
        \begin{tabular}{c}
                \includegraphics[width=7 cm]{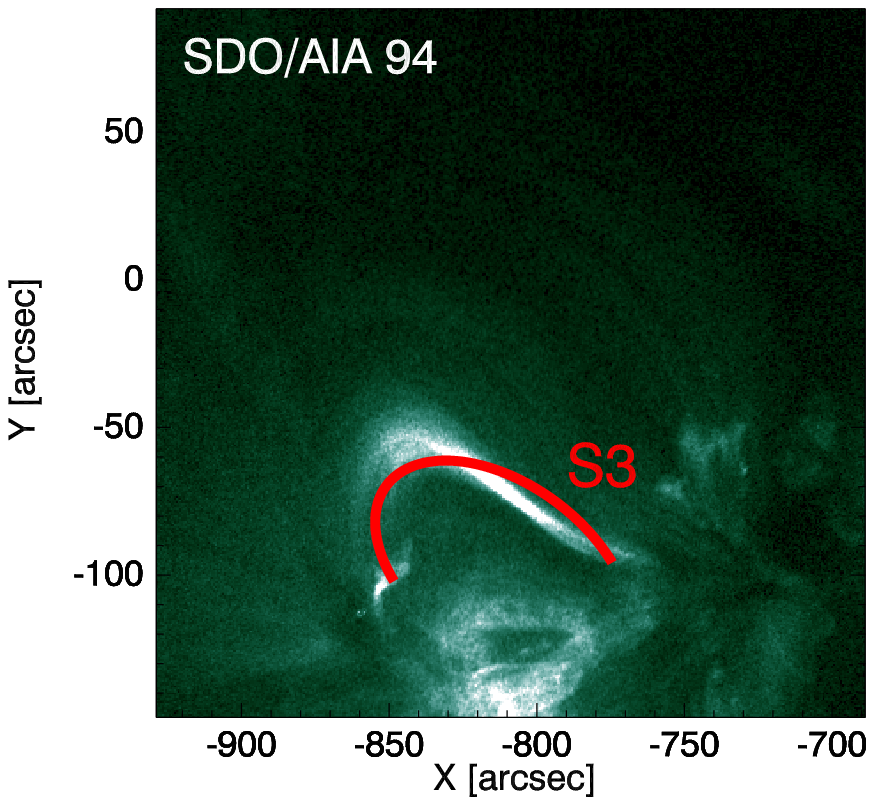} \\
                \includegraphics[width=7 cm]{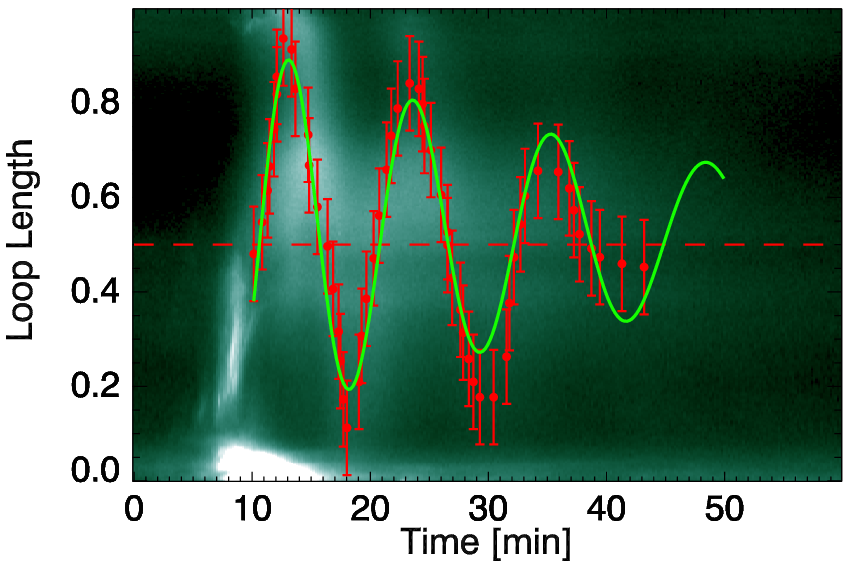} \\
                \includegraphics[width=7 cm]{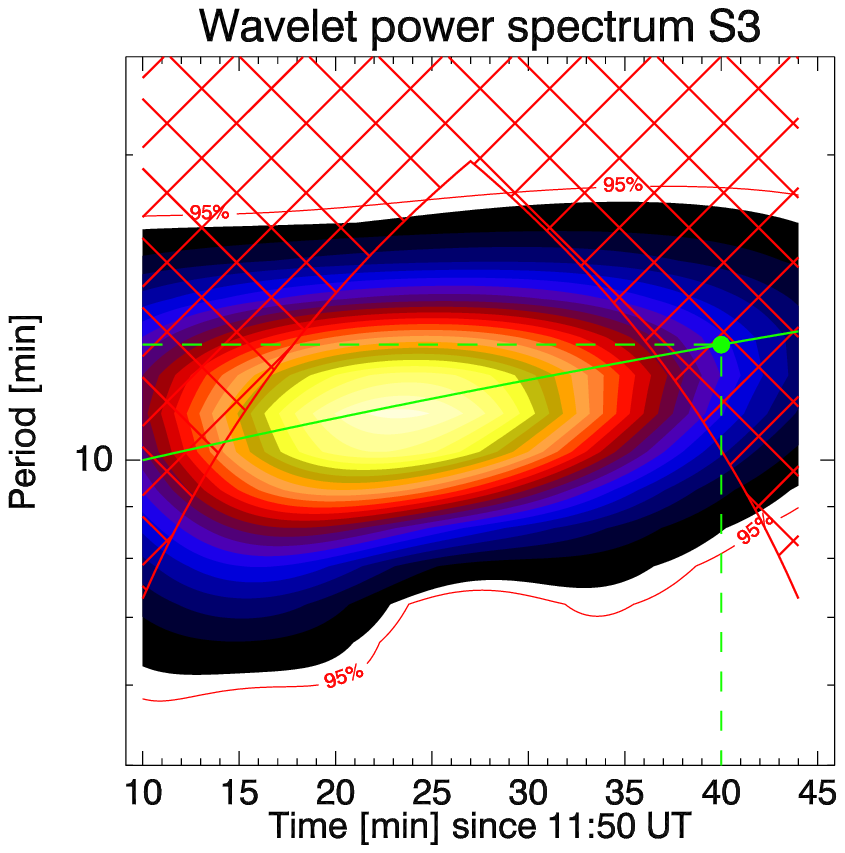} \\

        \end{tabular}
        \caption{Top: image of the loop at 94 \AA\ with the curved slit in red. Middle: TD maps from the curved slit. The red square points track the oscillation and are determined by eye. The oscillation in green is obtained by interpolation of the red points. Bottom: wavelet power spectrum of the oscillation profile. The green continuous line shows indicatively the rate at which the period of the slow wave varies (0.1 min min$^{-1}$), while the green dot marks the value of the period (13 min) after 30 min from its excitation.}
        \label{94img}
\end{figure}

We recall (see Fig.~\ref{temp_maps}) that the hot loop increases its brightness in the Be\_thin and Al\_poly filters 
and has a temperature of approximately 8 MK.
The high-cadence of the AIA instrument allows us to observe periodic intensity variations  along the hot loop in the 
94~\AA\ band.
 Indeed, after the flare, the brightness of the loop seems to vary periodically along the loop axis, bouncing between the footpoints \citep[similar to the observations reported in][]{Kumar2015}.
We have considered a curved slit along the loop (red curve in Fig. \ref{94img} - top) to extract the intensity along it from each frame and make a TD map (middle panel in Fig. \ref{94img}). An oscillatory pattern is clearly visible. 
To investigate the variation of the period, we have fitted the time series with Eq. \ref{eq_fitting}. The amplitude of the oscillation is almost half of the total loop length (0.4 $L$) and the initial period $P$ is approximately 10 min. The period variation per unit of time is $P' = 0.05 $ min min$^{-1}$, which is smaller than that estimated from the wavelet power spectrum (Fig. \ref{94img} - bottom). Indeed, the power spectrum exhibits a clear increase in the period over the time starting from 10 min, with an indicative rate of $P' = 0.1$ min min$^{-1}$ (continuous green line in the wavelet power spectrum).   

  \subsection{Coronal seismology with slow waves}

The longitudinal oscillation is essentially a slow magnetoacoustic wave, with a phase speed $V^{(\mathrm{S})}_\mathrm{ph}$
that can be determined from the loop length $L=L_{94} = 127\pm 13$ Mm, and the period $P= 9.7 \pm 0.4$ min:

\begin{equation}
V^{(\mathrm{S})}_\mathrm{ph} = \frac{2L}{P} = 436 \pm 47~\rm{km~s}^{-1}.
   \label{ph_speed_slow}
\end{equation}

In a low-$\beta$ plasma, the phase speed can be interpreted as the sound speed of a slow wave, whose magnitude depends upon the plasma temperature:

\begin{equation}
 C_\mathrm{S} = \left(\frac{\gamma p}{\rho}\right)^{1/2} = \left( \frac{2 \gamma k_\mathrm{B} T}{\mu m_p} \right)^{1/2}\approx \num{1.29e-1}(\gamma T/\mu)^{1/2}~\rm{km~s}^{-1}. 
\label{eq:eq_cs}
\end{equation}

If we consider $\gamma=5/3$, $\mu=1.27$ and an average temperature $T\sim8 \pm 2$ MK as determined by Hinode/XRT, the corresponding sound speed is $C_\mathrm{S} = 416 \pm 52~\rm{km~s}^{-1}$ (see also Eqs. \ref{app_cs_1}-\ref{app_cs_2}), which is in agreement with the estimated phase speed in Eq. \ref{ph_speed_slow}. 
However, the temperature estimates are affected by strong uncertainties ($\sim25\%$) (as well as the loop length in the absence of 3D stereoscopy) and the theoretical sound speed may be higher. In ideal MHD theory for cylindrical magnetic flux tubes, and accounting for finite-$\beta$ effects, the smallest phase speed for a slow MHD mode is given by the tube speed 
 
\begin{equation}
     C_\mathrm{T} = \frac{C_\mathrm{S} C_\mathrm{A}}{\sqrt{C_\mathrm{S}^2 + C_\mathrm{A}^2}}.
     \label{eq:tube_speed}
\end{equation}

For a temperature of $T=10 \pm 2.5$ MK, we have $C_\mathrm{S} = 465 \pm 58$ km s$^{-1}$, which is higher than the phase speed $V^{(\mathrm{S})}_\mathrm{ph}$. Therefore, assuming $C_\mathrm{T}=V^{(\mathrm{S})}_\mathrm{ph}$, we can determine the Alfv\'en speed for the hot loop \citep{Wang2007}. Indeed, from Eqs. \ref{eq:ca_csct}-\ref{eq:sigma_ca_csct}, and using the estimates for the sound and tube speeds, we find that $C_\mathrm{A} = 1255 \pm 207~\rm{km~s}^{-1}$.  
Assuming a typical magnetic field equal to that inferred from the kink oscillations, we can obtain an estimate for the density $n_0$ of the hot loop. Using the relations Eqs. \ref{eq:n0}-\ref{eq:sigma_n0}, we find $n_0=(6.0\pm3.0)\times10^9$ cm$^{-3}$, which is in agreement with the values found with Hinode/XRT (Fig. \ref{temp_maps} - bottom right panel). The values of the parameters from coronal seismology are summarised in Table \ref{table2}.

It is interesting to note that the variations over time of the period measured from the TD map and the temperature from Hinode/XRT data are consistent with each other. Indeed, it is easy to show that

\begin{equation}
\frac{\delta P}{P} = -\frac{\delta V^{(\mathrm{S})}_\mathrm{ph}}{V^{(\mathrm{S})}_\mathrm{ph}} ~~~{\rm and}~~~\frac{\delta V^{(\mathrm{S})}_\mathrm{ph}}{V^{(\mathrm{S})}_\mathrm{ph}} = \frac{\delta T}{2 T} ~~\rightarrow~~ \frac{\delta P}{P} = - \frac{\delta T}{2 T}
,\end{equation}
where $\delta P$ and $\delta T$ are the period and temperature variations in a given time interval, 
$P$ and $T$ being the initial values. 
We have not considered density variations, which may affect the phase speed via the dependence of the tube speed on the Alfv\'en speed. Moreover, the square root dependence of the Alfv\'en speed on the density decreases the effect of the density variation. However, the time profile of the density as inferred from Hinode/XRT is almost flat (see Fig. \ref{fig2}). 
In the range of 30 minutes, the period changes from 10 to 13 min (see the wavelet power spectrum in Fig. \ref{94img}),
 while the temperature drops from $\sim$8 MK to $\sim$3 MK (Fig. \ref{temp_maps}).
Therefore, having $\delta P = 3$ min and $\delta T$ = 5 MK, we find that $\delta P/P$ = 0.30 and $\delta T/2 T= 0.31$. 

Furthermore, by knowing the values of the period and plasma temperature, it is possible to inverse the problem in coronal seismology and estimate the length of the loop, for example, assuming that $V^{(\mathrm{S})}_\mathrm{ph} \approx C_\mathrm{S}$. For $P\sim10$ min and $T\sim8$ MK, we find $L \approx 125$ Mm, which is close to our estimate in Section \ref{sec_geo}. 

\section{Discussion and conclusion} 

Observations of fast and slow MHD modes in the same magnetic structure have a crucial role in coronal seismology. Indeed, in our analysis we show that the values of the magnetic field inferred from the observations of the kink oscillations and the coexisting longitudinal slow wave are in agreement. Therefore, it is possible to have a better understanding of the local environment and justify the robustness of this diagnostic technique \citep{Zhang2015}. In particular, from the knowledge of the local sound and Alfv\'en speeds, the value of the local plasma-$\beta$ is naturally ensued as,

\begin{equation}
 \beta = \frac{2}{\gamma} \left( \frac{C_\mathrm{S}}{C_\mathrm{A}}\right)^2.
 \label{eq:beta}
\end{equation}

By taking $C_\mathrm{S} =416-465~\rm{km~s}^{-1}$, $C_\mathrm{A} = 957-1255~\rm{km~s}^{-1}$ and $\gamma=5/3$, the plasma-$\beta$ will range between 0.14 and 0.28. On the other hand, the determination of these parameters presumes the knowledge of some other ones, whose values are usually assumed to be known by theory, such as the adiabatic index $\gamma$ and the mean molecular weight $\mu$. While the value of $\mu$ has less uncertainties and is assumed to the standard values of 1.27 from the  abundancy of ions in corona (hydrogen and helium), the local value of $\gamma$ is subject to discussion. The adiabatic index enters into the definition of the sound speed and determines the thermodynamics of plasma. Indeed, effective values of $\gamma$ may differ from the theoretical value of 5/3 because of partial ionisation, conductive and radiative cooling, or heating processes active in plasma. In our case, thermal conduction should not affect the dynamics of the hot loop since the the conduction time is estimated as \citep[p. 321 in ][]{Aschwanden2004},

\begin{equation}
 \tau_{cond} = \num{1.6e-9} n_0 T^{-5/2} L^2 \approx 142~\rm{min}.
\end{equation}       

In contrast, radiative processes may be important since the temperature of the hot loop decreases from $~$8 to $\sim$3 MK in $\sim$25 min, and the slow wave results to be over-damped (the damping time of the oscillation itself is 40 min  as determined by the fitting analysis). Therefore, values of $\gamma$ are assumed to vary between 1.1 and 5/3 \citep[p. 82 in][]{Priest2014}. \citet{VanDoorsselaere2011a} has provided the first measurement of the effective adiabatic index in the solar corona to be $\gamma_\mathrm{eff} = 1.10 \pm 0.02$, by analysing the density and temperature perturbations associated with a slow wave in ideal MHD theory approximation. However, this approach is valid when $\beta$ is very small \citep[see eq. (1) in][]{VanDoorsselaere2011a} and these perturbations are assumed to propagate like a pure sound wave, which is not the general case in corona.
Indeed, if we combine Eqs. \ref{eq:eq_cs}, \ref{eq:tube_speed} and \ref{eq:beta}, we can express the tube speed $C_\mathrm{T}$ in terms of $C_\mathrm{S}$ as

\begin{equation}
   C_\mathrm{T} = \frac{1}{\sqrt{1+\beta\gamma/2}}~C_\mathrm{S} = \left(\frac{\gamma}{1+\beta \gamma/2} \frac{2 k_B}{\mu m_p} T \right)^{1/2}.   \label{eq:ct_cs}
\end{equation}
Therefore, interpreting $C_\mathrm{T}$ as $C_\mathrm{S}$ can lead to erroneous results for $\gamma$ if the condition $\beta \ll 1$ is not satisfied, since it will be replaced by the factor  $\gamma_\mathrm{eff} = \gamma/(1+\beta \gamma /2)$, which will play the role of an effective adiabatic index. The relation between $\beta$, $\gamma$ and $\gamma_\mathrm{eff}$ can be written in the following form as,

\begin{equation}
 \beta = 2 \left( \frac{1}{\gamma_{\rm{eff}}} - \frac{1}{\gamma} \right).
 \label{beta_gammas}
\end{equation} 
Assuming a hypothetical value of $\gamma=5/3~(\approx $1.67 but in general it may be different), the corresponding value of $\gamma_\mathrm{eff}$ will decrease from 
5/3 according to the values assumed by $\beta$, which may vary through different coronal regions. It is trivial to show that for $\beta =0,$  $\gamma_\mathrm{eff} = \gamma$ and, hence, $C_\mathrm{T} = C_\mathrm{S}$. For $\beta$ = 0.01, which is a typical value assumed for active 
regions, we find $\gamma_\mathrm{eff}=1.65$ (in practise equal to $\gamma$; this justifies the low-$\beta$ approximation with the sound speed), but for values of $\beta \geq 0.13$, 
the difference becomes remarkable and $\gamma_\mathrm{eff}$ will deviate from the real and unknown $\gamma$ by more than 10 \%.
If we use $\gamma_\mathrm{eff}=1.1$ as estimated by \citet{VanDoorsselaere2011a}, we find $\beta\approx0.6$, which is unrealistic for that case: hence, the adiabatic index is surely much lower than 5/3.

\begin{figure}[htpb]
        \begin{tabular}{c}
                \includegraphics[width=9 cm]{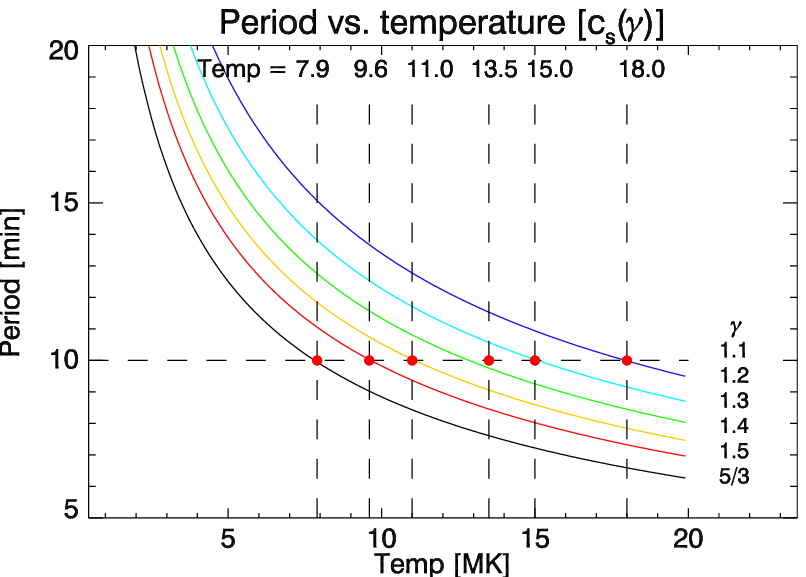} \\
                \includegraphics[width=9 cm]{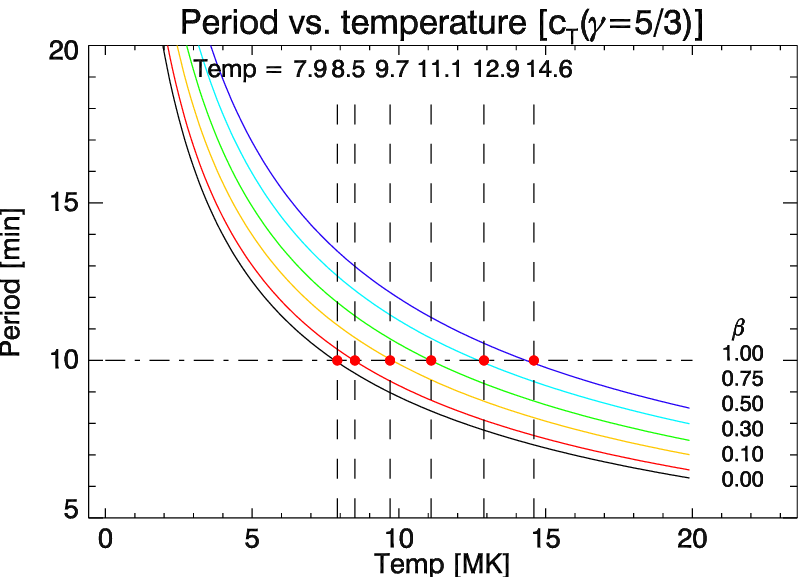} \\
                \includegraphics[width=9 cm]{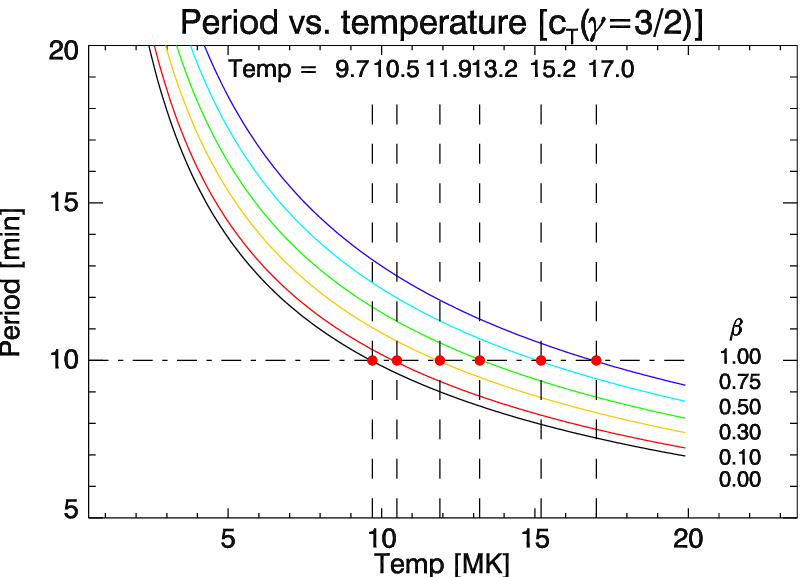} \\
        \end{tabular}
        \caption{Period vs. temperature for a slow wave, assuming phase speed equals the sound speed $C_\mathrm{S}$ (top) and the tube speed $C_\mathrm{T}$ (middle and bottom panels), respectively.}
        \label{period_vs_temp}
\end{figure}

The above discussion applied to our event can be easily illustrated by the plots of Fig. \ref{period_vs_temp}, which show how the period of a slow wave changes with the plasma temperature if we consider the phase speed to coincide with the sound (top panel) and the tube speeds (middle and bottom panels), respectively (the curves are obtained by combining Eqs. \ref{ph_speed_slow}--\ref{eq:eq_cs} and Eqs. \ref{ph_speed_slow}--\ref{eq:ct_cs}, and by taking the value of 127 Mm as the loop length). For the case of a pure sound wave, we plot different curves corresponding to different values of the adiabatic index $\gamma$ ranging between the theoretical value of 5/3 (lower curve in black) and 1.1 (upper curve in purple; the other values are indicated on the right of the plot). The intersection of the horizontal line at $P=$ 10 min (the initial period of our slow wave) with the different curves identifies the temperature at which the wave exists (for $\gamma=5/3,1.5,1.4,...$, $T_{\gamma} = 7.9, 9.6, 11.0, ...$ MK).  According to the analysis performed with XRT, the maximum temperature reached by the plasma is $\sim 10$ MK, and consequently the effective adiabatic index for our observation falls in the range $5/3 < \gamma \leq 1.5$. Similarly, we have also considered the tube speed for $\gamma = 5/3$ (middle panel) and $\gamma = 1.5$ (bottom panel) and plotted different curves for different values of the plasma-$\beta$ between 0.0 (lower curve in black) and 1.0 (upper curve in purple). For $\gamma=5/3$, the possible values of $\beta$ that fit our observations fall approximately in the range of 0.1 to 0.3, while for $\gamma = 1.5,$ the intersection points are moved towards higher temperatures and the lower limit at $T=10$ MK is obtained with $\beta = 0.0$, which again coincides with the case of a pure sound wave (but with $\gamma$=1.5). Therefore, based upon our results, we consider that the value for the adiabatic index $\gamma = 5/3$ and a plasma-$\beta$ between 0.1 and 0.3 can describe the dynamics of the slow MHD wave in the hot loop.

We would like to highlight that the correct interpretation of the nature of a slow MHD wave, and hence the density and temperature perturbations observed in the plasma, is essential for the correct inferences of the plasma parameters. We have shown that a finite value of the plasma-$\beta$, if not being sufficiently small, may lead to underestimated results for the adiabatic index $\gamma$. In general, coronal active regions are environments with a very small $\beta$ leading to experimental values $\gamma_\mathrm{eff}$ being very close to the real adiabatic index. However, this effect can be pronounced in the chromosphere and in the diffuse or higher corona at a distance greater than 2 R$_\odot$, where the approximation $\beta \ll 1$ is not valid, and must be taken into account in observations and numerical simulations. In this paper we have not considered other aspects perhaps relevant in the analysed event, for example a detailed study of the formation and dynamics of the cool ejection, the interaction with the coronal loop threads or the possible role of the observed MHD waves in heating. These could be the subject of future work.

\appendix
\section{Equations for coronal seismology}
\label{appendix}

General expression for the phase speed and its error for a fundamental standing MHD wave given the loop length $L\pm\sigma_L$ and the period $P\pm\sigma_P$: 
\begin{align}
 V_\mathrm{ph} & = \frac{2L}{P}, \label{app_ph_1}\\
\sigma_{V_\mathrm{ph}} & = V_\mathrm{ph}\sqrt{ (\sigma_{L}/L)^2 + (\sigma_{P}/P)^2 }. \label{app_ph_2} 
\end{align}

\subsection{Kink waves}

The Alfv\'en speed is determined by the kink speed $C_\mathrm{K} \pm \sigma_{C_\mathrm{K}}$ and the densities $n_e\pm \sigma_{n_e}$ and $n_0\pm\sigma_{n_0}$:
\begin{align}
C_\mathrm{A} & = C_\mathrm{K} \left( \frac{1+n_e/n_0}{2} \right)^{1/2}, \label{eq:ca}\\
\sigma_{C_\mathrm{A}} & = |C_\mathrm{A}| \sqrt{ \left(\frac{\sigma_{C_\mathrm{K}}}{C_\mathrm{K}}\right)^2 + \frac{(n_e/n_0)^2}{4(1+n_e/n_0)^2} \left[\left(\frac{\sigma_{n_e}}{n_e}\right)^2 + \left(\frac{\sigma_{n_0}}{n_0}\right)^2 \right]}. \label{eq:sigma_ca}
\end{align}

The magnetic field is given by the loop length, the period and internal and external densities with the associated uncertainties:
\begin{align}
B & = \frac{L}{P_\mathrm{K}} \sqrt{8\pi \mu m_p n_0 (1+n_e/n_0)}, \label{eq:b}\\
\sigma_B & = |B| \sqrt{ \left(\frac{\sigma_L}{L}\right)^2 + \left(\frac{\sigma_P}{P}\right)^2 + \frac{1}{4(n_0+n_e)^2}\left(\sigma_{n_e}^2 + \sigma_{n_0}^2 \right)}. \label{eq:sigma_b}
\end{align}

\subsection{Slow waves}
Expression for the sound speed given the plasma temperature $T\pm\sigma_T$:
\begin{align}
        C_\mathrm{S} & = \left( \frac{2 \gamma k_\mathrm{B} T}{\mu m_p} \right)^{1/2}, \label{app_cs_1}\\
        \sigma_{C_\mathrm{S}} & = \frac{1}{2} \left| C_\mathrm{S} \frac{\sigma_T}{T}\right|. \label{app_cs_2} 
\end{align}

The Alfv\'en speed is given by the tube ($C_\mathrm{T}$) and sound speeds:
\begin{align}
        C_\mathrm{A} & = \frac{ C_\mathrm{S} C_\mathrm{T}}{\sqrt{C_\mathrm{S}^2 - C_\mathrm{T}^2}}, \label{eq:ca_csct} \\
        \sigma_{C_\mathrm{A}} & = |C_\mathrm{A}| \sqrt{ \left(\frac{\sigma_{C_\mathrm{S}}}{C_\mathrm{S}}\right)^2 + \left(\frac{\sigma_{C_\mathrm{T}}}{C_\mathrm{T}}\right)^2 +\frac{1}{4}\left(\frac{\sigma_{C_\mathrm{S}^2-C_\mathrm{T}^2}}{C_\mathrm{S}^2-C_\mathrm{T}^2}\right)^2} ,\\
                              & \approx |C_\mathrm{A}| \sqrt{ \left(\frac{\sigma_{C_\mathrm{S}}}{C_\mathrm{S}}\right)^2 + \left(\frac{\sigma_{C_\mathrm{T}}}{C_\mathrm{T}}\right)^2 }. \label{eq:sigma_ca_csct}  
\end{align}
We note that in our analysis, the quantity $(C_\mathrm{S}^2 - C_\mathrm{T}^2)^{-2} \approx 10^{-9}$ km$^{-4}$ s$^4$, therefore it can be neglected.

The internal density $n_0$ is given by the magnetic field $B\pm\sigma_B$ and the Alfv\'en speed $C_\mathrm{A}\pm\sigma_{C_\mathrm{A}}$:
\begin{align}
         n_0 & = (4 \pi \mu m_p)^{-1} \left(\frac{B}{C_\mathrm{A}}\right)^2, \label{eq:n0}\\
\sigma_{n_0} & = 2|n_0| \sqrt{ \left( \frac{ \sigma_{C_\mathrm{A}}}{C_\mathrm{A}}\right)^2 + \left(\frac{\sigma_B}{B}\right)^2}. \label{eq:sigma_n0}
\end{align}

\begin{acknowledgements}
The present work was funded by STFC consolidated grant ST/L000733/1 (GN, VMN). VP acknowledges support from the Isaac Newton Studentship and the Cambridge Trust. The authors thank Helen Mason and the anonymous referee for their useful comments and suggestions. 
\end{acknowledgements}

\def\baselinestretch{1.0}
\bibliographystyle{aa}

\bibliography{references} 

\begin{thebibliography}{62}
\expandafter\ifx\csname natexlab\endcsname\relax\def\natexlab#1{#1}\fi

\bibitem[{{Anfinogentov} {et~al.}(2013){Anfinogentov}, {Nistic{\`o}}, \&
  {Nakariakov}}]{Anfinogentov2013}
{Anfinogentov}, S., {Nistic{\`o}}, G., \& {Nakariakov}, V.~M. 2013, \aap, 560,
  A107

\bibitem[{{Anfinogentov} {et~al.}(2015){Anfinogentov}, {Nakariakov}, \&
  {Nistic{\`o}}}]{Anfinogentov2015}
{Anfinogentov}, S.~A., {Nakariakov}, V.~M., \& {Nistic{\`o}}, G. 2015, \aap,
  583, A136

\bibitem[{{Aschwanden}(2004)}]{Aschwanden2004}
{Aschwanden}, M.~J. 2004, {Physics of the Solar Corona. An Introduction}
  (Praxis Publishing Ltd)

\bibitem[{{Aschwanden} {et~al.}(1999){Aschwanden}, {Fletcher}, {Schrijver}, \&
  {Alexander}}]{Aschwanden1999}
{Aschwanden}, M.~J., {Fletcher}, L., {Schrijver}, C.~J., \& {Alexander}, D.
  1999, \apj, 520, 880

\bibitem[{{Aschwanden} \& {Schrijver}(2011)}]{Aschwanden2011}
{Aschwanden}, M.~J. \& {Schrijver}, C.~J. 2011, \apj, 736, 102

\bibitem[{{Asplund} {et~al.}(2009){Asplund}, {Grevesse}, {Sauval}, \&
  {Scott}}]{Asplund09}
{Asplund}, M., {Grevesse}, N., {Sauval}, A.~J., \& {Scott}, P. 2009, \araa, 47,
  481

\bibitem[{{Brosius}(2013)}]{Brosius13}
{Brosius}, J.~W. 2013, \apj, 762, 133

\bibitem[{{Cooper} {et~al.}(2003){Cooper}, {Nakariakov}, \&
  {Tsiklauri}}]{Cooper2003}
{Cooper}, F.~C., {Nakariakov}, V.~M., \& {Tsiklauri}, D. 2003, \aap, 397, 765

\bibitem[{{Culhane} {et~al.}(2007){Culhane}, {Harra}, {James}, {Al-Janabi},
  {Bradley}, {Chaudry}, {Rees}, {Tandy}, {Thomas}, {Whillock}, {Winter},
  {Doschek}, {Korendyke}, {Brown}, {Myers}, {Mariska}, {Seely}, {Lang}, {Kent},
  {Shaughnessy}, {Young}, {Simnett}, {Castelli}, {Mahmoud}, {Mapson-Menard},
  {Probyn}, {Thomas}, {Davila}, {Dere}, {Windt}, {Shea}, {Hagood}, {Moye},
  {Hara}, {Watanabe}, {Matsuzaki}, {Kosugi}, {Hansteen}, \&
  {Wikstol}}]{Culhane97}
{Culhane}, J.~L., {Harra}, L.~K., {James}, A.~M., {et~al.} 2007, \solphys, 243,
  19

\bibitem[{{De Pontieu} {et~al.}(2014){De Pontieu}, {Title}, {Lemen}, {Kushner},
  {Akin}, {Allard}, {Berger}, {Boerner}, {Cheung}, {Chou}, {Drake}, {Duncan},
  {Freeland}, {Heyman}, {Hoffman}, {Hurlburt}, {Lindgren}, {Mathur}, {Rehse},
  {Sabolish}, {Seguin}, {Schrijver}, {Tarbell}, {W{\"u}lser}, {Wolfson},
  {Yanari}, {Mudge}, {Nguyen-Phuc}, {Timmons}, {van Bezooijen}, {Weingrod},
  {Brookner}, {Butcher}, {Dougherty}, {Eder}, {Knagenhjelm}, {Larsen},
  {Mansir}, {Phan}, {Boyle}, {Cheimets}, {DeLuca}, {Golub}, {Gates}, {Hertz},
  {McKillop}, {Park}, {Perry}, {Podgorski}, {Reeves}, {Saar}, {Testa}, {Tian},
  {Weber}, {Dunn}, {Eccles}, {Jaeggli}, {Kankelborg}, {Mashburn}, {Pust},
  {Springer}, {Carvalho}, {Kleint}, {Marmie}, {Mazmanian}, {Pereira}, {Sawyer},
  {Strong}, {Worden}, {Carlsson}, {Hansteen}, {Leenaarts}, {Wiesmann},
  {Aloise}, {Chu}, {Bush}, {Scherrer}, {Brekke}, {Martinez-Sykora}, {Lites},
  {McIntosh}, {Uitenbroek}, {Okamoto}, {Gummin}, {Auker}, {Jerram}, {Pool}, \&
  {Waltham}}]{DePontieu14}
{De Pontieu}, B., {Title}, A.~M., {Lemen}, J.~R., {et~al.} 2014, \solphys, 289,
  2733

\bibitem[{{DeForest} \& {Gurman}(1998)}]{DeForest1998}
{DeForest}, C.~E. \& {Gurman}, J.~B. 1998, \apjl, 501, L217

\bibitem[{{Del Zanna}(2013{\natexlab{a}})}]{DelZanna13}
{Del Zanna}, G. 2013{\natexlab{a}}, \aap, 555, A47

\bibitem[{{Del Zanna}(2013{\natexlab{b}})}]{delzanna:2013_multithermal}
{Del Zanna}, G. 2013{\natexlab{b}}, \aap, 558, A73

\bibitem[{{Del Zanna} {et~al.}(2006){Del Zanna}, {Berlicki}, {Schmieder}, \&
  {Mason}}]{DelZanna06}
{Del Zanna}, G., {Berlicki}, A., {Schmieder}, B., \& {Mason}, H.~E. 2006,
  \solphys, 234, 95

\bibitem[{{Del Zanna} {et~al.}(2015){Del Zanna}, {Dere}, {Young}, {Landi}, \&
  {Mason}}]{delzanna_etal:2015_chianti_v8}
{Del Zanna}, G., {Dere}, K.~P., {Young}, P.~R., {Landi}, E., \& {Mason}, H.~E.
  2015, \aap, 582, A56

\bibitem[{{Del Zanna} {et~al.}(2011){Del Zanna}, {O'Dwyer}, \&
  {Mason}}]{delzanna_etal:11_aia}
{Del Zanna}, G., {O'Dwyer}, B., \& {Mason}, H.~E. 2011, \aap, 535, A46

\bibitem[{{Dere} {et~al.}(2009){Dere}, {Landi}, {Young}, {Del Zanna},
  {Landini}, \& {Mason}}]{Dere2009}
{Dere}, K.~P., {Landi}, E., {Young}, P.~R., {et~al.} 2009, \aap, 498, 915

\bibitem[{{Edwin} \& {Roberts}(1983)}]{Edwin1983}
{Edwin}, P.~M. \& {Roberts}, B. 1983, \solphys, 88, 179

\bibitem[{{Goddard} {et~al.}(2016){Goddard}, {Nistic{\`o}}, {Nakariakov}, \&
  {Zimovets}}]{Goddard2016}
{Goddard}, C.~R., {Nistic{\`o}}, G., {Nakariakov}, V.~M., \& {Zimovets}, I.~V.
  2016, \aap, 585, A137

\bibitem[{{Golub} {et~al.}(2007){Golub}, {Deluca}, {Austin}, {Bookbinder},
  {Caldwell}, {Cheimets}, {Cirtain}, {Cosmo}, {Reid}, {Sette}, {Weber},
  {Sakao}, {Kano}, {Shibasaki}, {Hara}, {Tsuneta}, {Kumagai}, {Tamura},
  {Shimojo}, {McCracken}, {Carpenter}, {Haight}, {Siler}, {Wright}, {Tucker},
  {Rutledge}, {Barbera}, {Peres}, \& {Varisco}}]{Golub2007}
{Golub}, L., {Deluca}, E., {Austin}, G., {et~al.} 2007, \solphys, 243, 63

\bibitem[{{Hershaw} {et~al.}(2011){Hershaw}, {Foullon}, {Nakariakov}, \&
  {Verwichte}}]{Hershaw2011}
{Hershaw}, J., {Foullon}, C., {Nakariakov}, V.~M., \& {Verwichte}, E. 2011,
  \aap, 531, A53

\bibitem[{{Hood} {et~al.}(2013){Hood}, {Ruderman}, {Pascoe}, {De Moortel},
  {Terradas}, \& {Wright}}]{Hood2013}
{Hood}, A.~W., {Ruderman}, M., {Pascoe}, D.~J., {et~al.} 2013, \aap, 551, A39

\bibitem[{{Kiddie} {et~al.}(2012){Kiddie}, {De Moortel}, {Del Zanna},
  {McIntosh}, \& {Whittaker}}]{Kiddie2012}
{Kiddie}, G., {De Moortel}, I., {Del Zanna}, G., {McIntosh}, S.~W., \&
  {Whittaker}, I. 2012, \solphys, 279, 427

\bibitem[{{Kim} {et~al.}(2012){Kim}, {Nakariakov}, \& {Shibasaki}}]{Kim2012}
{Kim}, S., {Nakariakov}, V.~M., \& {Shibasaki}, K. 2012, \apjl, 756, L36

\bibitem[{{Kobelski} {et~al.}(2014){Kobelski}, {Saar}, {Weber}, {McKenzie}, \&
  {Reeves}}]{Kobelski2014}
{Kobelski}, A.~R., {Saar}, S.~H., {Weber}, M.~A., {McKenzie}, D.~E., \&
  {Reeves}, K.~K. 2014, \solphys, 289, 2781

\bibitem[{{Krishna Prasad} {et~al.}(2012){Krishna Prasad}, {Banerjee}, {Van
  Doorsselaere}, \& {Singh}}]{Prasad2012}
{Krishna Prasad}, S., {Banerjee}, D., {Van Doorsselaere}, T., \& {Singh}, J.
  2012, \aap, 546, A50

\bibitem[{{Kumar} {et~al.}(2015){Kumar}, {Nakariakov}, \& {Cho}}]{Kumar2015}
{Kumar}, P., {Nakariakov}, V.~M., \& {Cho}, K.-S. 2015, \apj, 804, 4

\bibitem[{{Kupriyanova} {et~al.}(2013){Kupriyanova}, {Melnikov}, \&
  {Shibasaki}}]{Kupriyanova2013}
{Kupriyanova}, E.~G., {Melnikov}, V.~F., \& {Shibasaki}, K. 2013, \pasj, 65

\bibitem[{{Lemen} {et~al.}(2012){Lemen}, {Title}, {Akin}, {Boerner}, {Chou},
  {Drake}, {Duncan}, {Edwards}, {Friedlaender}, {Heyman}, {Hurlburt}, {Katz},
  {Kushner}, {Levay}, {Lindgren}, {Mathur}, {McFeaters}, {Mitchell}, {Rehse},
  {Schrijver}, {Springer}, {Stern}, {Tarbell}, {Wuelser}, {Wolfson}, {Yanari},
  {Bookbinder}, {Cheimets}, {Caldwell}, {Deluca}, {Gates}, {Golub}, {Park},
  {Podgorski}, {Bush}, {Scherrer}, {Gummin}, {Smith}, {Auker}, {Jerram},
  {Pool}, {Soufli}, {Windt}, {Beardsley}, {Clapp}, {Lang}, \&
  {Waltham}}]{Lemen2012}
{Lemen}, J.~R., {Title}, A.~M., {Akin}, D.~J., {et~al.} 2012, \solphys, 275, 17

\bibitem[{{Liu} {et~al.}(2012){Liu}, {Ofman}, {Nitta}, {Aschwanden},
  {Schrijver}, {Title}, \& {Tarbell}}]{Liu2012}
{Liu}, W., {Ofman}, L., {Nitta}, N.~V., {et~al.} 2012, \apj, 753, 52

\bibitem[{{Markwardt}(2009)}]{Markwardt2009}
{Markwardt}, C.~B. 2009, in Astronomical Society of the Pacific Conference
  Series, Vol. 411, Astronomical Data Analysis Software and Systems XVIII, ed.
  D.~A. {Bohlender}, D.~{Durand}, \& P.~{Dowler}, 251

\bibitem[{{Nakariakov} {et~al.}(1999){Nakariakov}, {Ofman}, {Deluca},
  {Roberts}, \& {Davila}}]{Nakariakov1999}
{Nakariakov}, V.~M., {Ofman}, L., {Deluca}, E.~E., {Roberts}, B., \& {Davila},
  J.~M. 1999, Science, 285, 862

\bibitem[{{Nakariakov} {et~al.}(2016){Nakariakov}, {Pilipenko}, {Heilig},
  {Jel{\'{\i}}nek}, {Karlick{\'y}}, {Klimushkin}, {Kolotkov}, {Lee},
  {Nistic{\`o}}, {Van Doorsselaere}, {Verth}, \& {Zimovets}}]{Nakariakov2016}
{Nakariakov}, V.~M., {Pilipenko}, V., {Heilig}, B., {et~al.} 2016, \ssr, 200,
  75

\bibitem[{{Narukage} {et~al.}(2014){Narukage}, {Sakao}, {Kano}, {Shimojo},
  {Winebarger}, {Weber}, \& {Reeves}}]{Narukage2014}
{Narukage}, N., {Sakao}, T., {Kano}, R., {et~al.} 2014, \solphys, 289, 1029

\bibitem[{{Nistic{\`o}} {et~al.}(2013{\natexlab{a}}){Nistic{\`o}},
  {Nakariakov}, \& {Verwichte}}]{Nistico2013}
{Nistic{\`o}}, G., {Nakariakov}, V.~M., \& {Verwichte}, E. 2013{\natexlab{a}},
  \aap, 552, A57

\bibitem[{{Nistic{\`o}} {et~al.}(2014){Nistic{\`o}}, {Pascoe}, \&
  {Nakariakov}}]{Nistico2014}
{Nistic{\`o}}, G., {Pascoe}, D.~J., \& {Nakariakov}, V.~M. 2014, \aap, 569, A12

\bibitem[{{Nistic{\`o}} {et~al.}(2013{\natexlab{b}}){Nistic{\`o}}, {Verwichte},
  \& {Nakariakov}}]{Nistico3d}
{Nistic{\`o}}, G., {Verwichte}, E., \& {Nakariakov}, V. 2013{\natexlab{b}},
  Entropy, 15, 4520

\bibitem[{{O'Dwyer} {et~al.}(2014{\natexlab{a}}){O'Dwyer}, {Del Zanna}, \&
  {Mason}}]{odwyer_etal:2014}
{O'Dwyer}, B., {Del Zanna}, G., \& {Mason}, H.~E. 2014{\natexlab{a}}, \aap,
  561, A20

\bibitem[{{O'Dwyer} {et~al.}(2014{\natexlab{b}}){O'Dwyer}, {Del Zanna}, \&
  {Mason}}]{ODwyer2014}
{O'Dwyer}, B., {Del Zanna}, G., \& {Mason}, H.~E. 2014{\natexlab{b}}, \aap,
  561, A20

\bibitem[{{O'Dwyer} {et~al.}(2010){O'Dwyer}, {Del Zanna}, {Mason}, {Weber}, \&
  {Tripathi}}]{odwyer_etal:10}
{O'Dwyer}, B., {Del Zanna}, G., {Mason}, H.~E., {Weber}, M.~A., \& {Tripathi},
  D. 2010, \aap, 521, A21+

\bibitem[{{Oliver} {et~al.}(2016){Oliver}, {Soler}, {Terradas}, \&
  {Zaqarashvili}}]{Oliver2016}
{Oliver}, R., {Soler}, R., {Terradas}, J., \& {Zaqarashvili}, T.~V. 2016, \apj,
  818, 128

\bibitem[{{Oliver} {et~al.}(2014){Oliver}, {Soler}, {Terradas}, {Zaqarashvili},
  \& {Khodachenko}}]{Oliver2014}
{Oliver}, R., {Soler}, R., {Terradas}, J., {Zaqarashvili}, T.~V., \&
  {Khodachenko}, M.~L. 2014, \apj, 784, 21

\bibitem[{{Pascoe} {et~al.}(2016){Pascoe}, {Goddard}, {Nistic{\`o}},
  {Anfinogentov}, \& {Nakariakov}}]{Pascoe2016}
{Pascoe}, D.~J., {Goddard}, C.~R., {Nistic{\`o}}, G., {Anfinogentov}, S., \&
  {Nakariakov}, V.~M. 2016, \aap, 585, L6

\bibitem[{{Patsourakos} \& {Vourlidas}(2012)}]{Patsourakos2012}
{Patsourakos}, S. \& {Vourlidas}, A. 2012, \solphys, 281, 187

\bibitem[{{Petkaki} {et~al.}(2012){Petkaki}, {Del Zanna}, {Mason}, \&
  {Bradshaw}}]{petkaki_etal:12}
{Petkaki}, P., {Del Zanna}, G., {Mason}, H.~E., \& {Bradshaw}, S. 2012, \aap,
  547, A25

\bibitem[{{Priest}(2014)}]{Priest2014}
{Priest}, E. 2014, {Magnetohydrodynamics of the Sun} (Cambridge University
  Press)

\bibitem[{{Roberts} {et~al.}(1984){Roberts}, {Edwin}, \& {Benz}}]{Roberts1984}
{Roberts}, B., {Edwin}, P.~M., \& {Benz}, A.~O. 1984, \apj, 279, 857

\bibitem[{{Terradas} \& {Ofman}(2004)}]{Terradas2004}
{Terradas}, J. \& {Ofman}, L. 2004, \apj, 610, 523

\bibitem[{{Thurgood} \& {McLaughlin}(2013)}]{Thurgood2013}
{Thurgood}, J.~O. \& {McLaughlin}, J.~A. 2013, \solphys, 288, 205

\bibitem[{{Tikhonchuk} {et~al.}(1995){Tikhonchuk}, {Rankin}, {Frycz}, \&
  {Samson}}]{Tikhonchuk1995}
{Tikhonchuk}, V.~T., {Rankin}, R., {Frycz}, P., \& {Samson}, J.~C. 1995,
  Physics of Plasmas, 2, 501

\bibitem[{{Van Doorsselaere} {et~al.}(2011{\natexlab{a}}){Van Doorsselaere},
  {De Groof}, {Zender}, {Berghmans}, \& {Goossens}}]{VanDoorsselaere2011b}
{Van Doorsselaere}, T., {De Groof}, A., {Zender}, J., {Berghmans}, D., \&
  {Goossens}, M. 2011{\natexlab{a}}, \apj, 740, 90

\bibitem[{{Van Doorsselaere} {et~al.}(2011{\natexlab{b}}){Van Doorsselaere},
  {Wardle}, {Del Zanna}, {Jansari}, {Verwichte}, \&
  {Nakariakov}}]{VanDoorsselaere2011a}
{Van Doorsselaere}, T., {Wardle}, N., {Del Zanna}, G., {et~al.}
  2011{\natexlab{b}}, \apjl, 727, L32

\bibitem[{{Verwichte} {et~al.}(2009){Verwichte}, {Aschwanden}, {Van
  Doorsselaere}, {Foullon}, \& {Nakariakov}}]{Verwichte2009}
{Verwichte}, E., {Aschwanden}, M.~J., {Van Doorsselaere}, T., {Foullon}, C., \&
  {Nakariakov}, V.~M. 2009, \apj, 698, 397

\bibitem[{{Verwichte} {et~al.}(2010){Verwichte}, {Foullon}, \& {Van
  Doorsselaere}}]{Verwichte2010}
{Verwichte}, E., {Foullon}, C., \& {Van Doorsselaere}, T. 2010, \apj, 717, 458

\bibitem[{{Verwichte} {et~al.}(1999){Verwichte}, {Nakariakov}, \&
  {Longbottom}}]{Verwichte1999}
{Verwichte}, E., {Nakariakov}, V.~M., \& {Longbottom}, A.~W. 1999, Journal of
  Plasma Physics, 62, 219

\bibitem[{{Verwichte} {et~al.}(2013){Verwichte}, {Van Doorsselaere}, {Foullon},
  \& {White}}]{Verwichte2013}
{Verwichte}, E., {Van Doorsselaere}, T., {Foullon}, C., \& {White}, R.~S. 2013,
  \apj, 767, 16

\bibitem[{{Wang} {et~al.}(2007){Wang}, {Innes}, \& {Qiu}}]{Wang2007}
{Wang}, T., {Innes}, D.~E., \& {Qiu}, J. 2007, \apj, 656, 598

\bibitem[{{Wang} {et~al.}(2012){Wang}, {Ofman}, {Davila}, \& {Su}}]{Wang2012}
{Wang}, T., {Ofman}, L., {Davila}, J.~M., \& {Su}, Y. 2012, \apjl, 751, L27

\bibitem[{{Yuan} \& {Van Doorsselaere}(2016{\natexlab{a}})}]{Ding2016a}
{Yuan}, D. \& {Van Doorsselaere}, T. 2016{\natexlab{a}}, \apjs, 223, 23

\bibitem[{{Yuan} \& {Van Doorsselaere}(2016{\natexlab{b}})}]{Ding2016b}
{Yuan}, D. \& {Van Doorsselaere}, T. 2016{\natexlab{b}}, \apjs, 223, 24

\bibitem[{{Zhang} {et~al.}(2015){Zhang}, {Zhang}, {Wang}, \&
  {Nakariakov}}]{Zhang2015}
{Zhang}, Y., {Zhang}, J., {Wang}, J., \& {Nakariakov}, V.~M. 2015, \aap, 581,
  A78

\bibitem[{{Zimovets} \& {Nakariakov}(2015)}]{Zimovets2015}
{Zimovets}, I.~V. \& {Nakariakov}, V.~M. 2015, \aap, 577, A4

\end{thebibliography}
\end{document}